\documentclass[amsmath,amssymb,reprint,superscriptaddress,floatfix, nofootinbib]{revtex4-2}
\usepackage[dvipsnames]{xcolor}
\usepackage{graphicx}
\usepackage{changepage} 
\usepackage{multirow} 
\usepackage{hyperref}
\usepackage{soul} 
\usepackage[maxfloats=256]{morefloats}
\usepackage{ulem}
\usepackage{booktabs}
\usepackage{url}
\usepackage[percent]{overpic} 
\usepackage{bbm}
\usepackage{xurl}
\usepackage[capitalise,noabbrev,nameinlink]{cleveref}
\hypersetup{
citecolor=blue,
 colorlinks=true,
 linkcolor=blue,
 filecolor=magenta,
 urlcolor=blue,
 breaklinks=true,
}

\usepackage{orcidlink}
\graphicspath{{./images/}}
\Crefname{equation}{Eq.}{Eqs.}

\newcommand{\ch}[1]{\chi_{#1}}

\newcommand{\cmfpp}{\texttt{CMF++}}
\newcommand{\eos}{EOS}
\newcommand{\bvarphi}{\boldsymbol{\varphi}}

\begin{document}

\title{Strangeness in the Chiral Mean Field model phase diagram}
\author{Mateus Reinke Pelicer\,\orcidlink{0000-0002-2189-706X}}
\affiliation{The Grainger College of Engineering, Illinois Center for Advanced Studies of the Universe, Department of Physics, University of Illinois at Urbana-Champaign, Urbana, IL 61801, USA}
\author{Nikolas Cruz-Camacho\,\orcidlink{0009-0004-7870-0039}}
\affiliation{The Grainger College of Engineering, Illinois Center for Advanced Studies of the Universe, Department of Physics, University of Illinois at Urbana-Champaign, Urbana, IL 61801, USA}
\author{Rajesh Kumar\,\orcidlink{0000-0003-2746-3956}}
\affiliation{Department of Physics, Kent State University, Kent, OH 44243, USA}
\affiliation{Department of Physics, MRPD Government College Talwara, Punjab, 144216, India}
\author{Veronica Dexheimer\,\orcidlink{0000-0001-5578-2626}}
\affiliation{Department of Physics, Kent State University, Kent, OH 44243, USA}
\affiliation{CERCA/ISO, Department of Physics, Case Western Reserve University, Cleveland, Ohio 44106, USA}
\author{Jacquelyn Noronha-Hostler\,\orcidlink{0000-0003-3229-4958}}
\affiliation{The Grainger College of Engineering, Illinois Center for Advanced Studies of the Universe, Department of Physics, University of Illinois at Urbana-Champaign, Urbana, IL 61801, USA}

\collaboration{MUSES Collaboration}
\noaffiliation
\date{\today}

\begin{abstract}
{We explore the phase diagram of the Chiral Mean Field model in the space of temperature and baryon, charge, and strangeness chemical potentials, focusing on the role of the strangeness direction.
At low density, a negative $\mu_S$ leads to finite strangeness in the nuclear liquid--gas transition and rotates the order parameter away from the purely baryonic direction. The critical mode remains baryon dominated over the $\mu_S$ range considered, while the critical locus develops a strong parametrization dependence once multistrange baryons appear. At higher density, a separate first-order transition toward strangeness-rich hadronic matter appears only for one of the two parametrizations considered, emphasizing its interaction dependence. The high density transition reorganizes octet-dominated matter into decuplet-dominated matter: $\Delta$ baryons provide most of the baryon-density discontinuity, while $\Xi^*(1530)$ baryons provide the largest contribution to the strangeness-density discontinuity. Its order parameter is dominated by the strangeness density, and its normalized direction in the $BQS$ space changes little with temperature even as the discontinuity vanishes at the critical point. The critical baryon chemical potential changes by almost $1$~GeV over $\mu_S\in [-300, 300]$~MeV. We characterize both transitions using the fixed-temperature $BQS$ susceptibility matrix, the density discontinuity between the coexisting phases, and a generalized Clausius--Clapeyron relation.
This work also accompanies the release of \cmfpp{} \texttt{v2.0.0} as part of the MUSES framework. The extension of the initial release includes, among other features, finite-temperature thermodynamics, new interactions and parametrizations, and  susceptibility calculations based on a Jacobian method.

}
\end{abstract}

\maketitle

\section{Introduction}
\label{sec:intro}

The thermodynamics of strongly interacting matter is organized by the charges conserved by the strong interaction. Relevant for our discussion are the baryon number $B$, electric charge $Q$, and strangeness $S$. The equation of state (\eos) is therefore a function on the four-dimensional space of temperature and three chemical potentials, $(T,\mu_B,\mu_Q,\mu_S)$.
Based on these conserved quantities, the individual particle chemical potentials are
\begin{equation}
\mu_i=B_i\mu_B+Q_i\mu_Q+S_i\mu_S,
\label{eq:mu_bare}
\end{equation}
with strange baryons carrying $S_i<0$.

Different physical systems---heavy-ion collisions, neutron stars and post-merger remnants, nuclei, and hypernuclei---explore a constrained subspace of this thermodynamic domain. Some of these constraints follow from global conservation laws, while others reflect assumptions of local neutrality or chemical equilibrium. Over finite length and time scales, however, the corresponding densities can fluctuate and chemical equilibrium may not be maintained. Therefore, exploring the chemical potentials independently can reveal thermodynamic responses and phase structures that remain hidden along the constrained trajectories. The systems in which the strange direction can become relevant are discussed in~\Cref{sec:strangeness}.

At supra-saturation density, strangeness can generate a qualitatively distinct phase structure. While hyperons often appear continuously as their chemical-potential thresholds are crossed, sufficiently attractive hyperon--hyperon interactions can produce a separate minimum of the thermodynamic potential. Depending on the assumed interaction, this may lead to a first-order transition between phases with different hyperonic compositions~\cite{Schaffner-Bielich:2000igu,Schaffner-Bielich:2000nft}. Subsequent thermodynamic studies of $n\Lambda$, $np\Lambda$, and full-octet matter formulated these transitions in terms of the convexity of the free energy and the associated multidimensional Gibbs construction~\cite{Gulminelli:2012iq,Gulminelli:2013qr,Raduta:2014lja}.  The existence of these strangeness related phase transitions is strongly interaction dependent but, so far, constrained relativistic mean-field calculations have found no supra-saturation instability within the parameter space compatible with modern hypernuclear potentials~\cite{Torres:2016ydl}. Nevertheless, the treatment of the baryon decuplet and its mass generation introduces an additional source of uncertainty that can enhance the transition~\cite{Zschiesche:2000ew}. Such a novel phase would have potentially important consequences for neutron star composition and dynamics.

The chiral mean field (CMF) model provides a natural framework for studying strangeness. It is based on a nonlinear realization of chiral ${\rm SU}(3)$ symmetry and describes the baryon octet and decuplet, in addition to up, down, and strange quarks, coupled to the non-strange and strange scalar condensates. Previous CMF studies have shown that the strange sector can qualitatively modify the compact-star phase structure~\cite{Dexheimer:2014pea}. Our previous zero-temperature analysis, in which $\mu_B$, $\mu_Q$, and $\mu_S$ were varied independently, found a high-density hadronic first-order transition in some CMF parametrizations but not in others~\cite{Cruz-Camacho:2024odu}. Here, we ask how a finite $\mu_S$ reorganizes the nuclear liquid--gas transition and how the coexistence surface and critical locus (the collection of critical points) of the high-density transition evolve at finite temperature. To study stability in this multidimensional space, we calculate the complete second-order susceptibility matrix using the Jacobian method of Ref.~\cite{Gholami:2025cfq}.

Focusing on the redefined RC2 and RC4 representative parametrizations of CMF introduced in~\cite{Kumar:2024owe}, we find that decreasing $\mu_S$ introduces finite strangeness into the nuclear liquid--gas transition and changes both the direction of its order parameter in the $BQS$ space and the trajectory of its critical point in $\mu_B$ and $T$ space. At higher density, RC2 exhibits a separate first-order hadronic transition that persists at finite temperature, while no corresponding transition is found with RC4. This transition reorganizes octet-dominated matter into decuplet-dominated matter. Its density discontinuity is strangeness dominated and retains a nearly temperature-independent direction as the critical point is approached. These calculations accompany the open-source release of \cmfpp{} \texttt{v2.0.0}, developed as part of the MUSES collaboration~\cite{ReinkePelicer:2025vuh,Jahan:2026hvs}. The release makes the finite-temperature thermodynamics, Jacobian-based susceptibilities, and numerical phase-search methods used here available both as a stand-alone code and through the MUSES Calculation Engine~\cite{pelicer_2026_24091453,muses_calculation_engine_2025_v1}.

The paper is organized as follows. In~\Cref{sec:strangeness} we discuss the physical systems and thermodynamic constraints that motivate exploring the strange chemical direction. In~\Cref{sec:formalism} we present the finite-temperature CMF model, the available parametrizations and particle contents, the Jacobian-based susceptibility calculation and stability classification, and the numerical implementation. In~\Cref{sec:results} we validate the susceptibility calculation, then we analyze the liquid--gas transition and the high-density strangeness-dominated transition. \Cref{sec:conclusions} summarizes our conclusions. Details of the decuplet couplings are collected in the Appendix.

We work in natural units, $\hbar=c=k_B=1$. 

\section{Physical systems and the strangeness direction}
\label{sec:strangeness}

Different physical systems correspond to different trajectories through the thermodynamic space $(T,\mu_B,\mu_Q,\mu_S)$. These trajectories are commonly defined by imposing charge neutrality, fixed charge fractions, or chemical-equilibrium conditions. Although these constraints are appropriate for constructing equilibrium equations of state, they do not remove the underlying conserved-charge directions. The corresponding densities can fluctuate locally, and chemical equilibrium may not be maintained when the system evolves faster than the relevant reactions equilibrate.

\cmfpp{} can be used to calculate thermodynamic properties and particle populations across the QCD phase diagram. Because it includes both hadronic and quark degrees of freedom, it can describe different phases in different thermodynamic regimes. Here, we summarize the systems in which strangeness is relevant, the constraints imposed by their conserved quantities, and the regions of thermodynamic space that they explore.

In cold neutron stars, the temperature can usually be neglected, and weak reactions have sufficient time to establish chemical equilibrium. Since strangeness is not conserved by the weak interaction, this gives $\mu_S=0$. Charge neutrality then fixes $\mu_Q$ at each $\mu_B$, reducing cold catalyzed matter to a one-dimensional trajectory that can be parametrized by either $\mu_B$ or $n_B$~\cite{Glendenning:1991es,Schaffner:1995th,Chatterjee:2015pua}. This condition depends, however, on the weak reactions being fast enough to follow the evolution of the system. Non-leptonic reactions such as $nn\leftrightarrow n\Lambda$ and hyperonic Urca processes operate on finite timescales~\cite{vanDalen:2003uy,Gusakov:2008hv,Ofengeim:2019fjy,Alford:2020pld,Ghosh:2025glz}. During binary inspiral or stellar oscillations, density perturbations can drive matter away from weak equilibrium. A recent reaction-network calculation using a four-dimensional \eos{} with strangeness found that perturbations at the low temperatures relevant for the inspiral can generate both $\mu_S\neq0$ and a nonzero beta-equilibrium imbalance $\mu_\beta\equiv\mu_n-\mu_p-\mu_e$~\cite{Storbacka:2026ehf}.

In heavy-ion collisions the incoming nuclei carry zero net strangeness and an electric charge fraction fixed by the colliding species. Equilibrium equations of state used in hydrodynamic simulations therefore commonly impose $n_S=0$ and a fixed $Y_Q=n_Q/n_B$, which determine $\mu_S$ and $\mu_Q$ as functions of $(T,\mu_B)$~\cite{Bazavov:2012vg,Fu:2018qsk,Noronha-Hostler:2019ayj,Lysenko:2024hqp}. For ideal hydrodynamic flow, conservation of entropy and baryon number further implies each fluid element follows an isentropic trajectory, which sets $T(\mu_B)$. But strangeness neutrality does not imply the absence of strange particles, and at finite $\mu_B$, strange baryons and particles carrying the opposite strangeness can remain abundant while their net contributions cancel. Moreover, only the total strangeness of the collision system is fixed. Local fluctuations of the conserved-charge densities are allowed and are characterized in equilibrium by the generalized susceptibilities of the QCD equation of state~\cite{Jeon:2003gk,Koch:2005vg,HotQCD:2012fhj,Bazavov:2012vg,Bellwied:2013cta,Borsanyi:2014ewa}. Experimentally, related correlations can be studied through event-by-event fluctuations of identified hadron multiplicities~
\cite{STAR:2017tfy,STAR:2013gus,STAR:2014egu,STAR:2020tga,STAR:2020ddh,ALICE:2022xpf}.  
Recently, a recent Bayesian analysis of hadron yields from STAR Ru+Ru and Zr+Zr isobar collisions extracted differences among $\mu_B$, $\mu_Q$, and $\mu_S$ and their ratios. The results demonstrate that controlled variations of the charge fraction can probe correlated directions in the four-dimensional QCD thermodynamic space~\cite{Grefa:2026meq}.
Varying the chemical potentials independently is therefore not meant to represent a single equilibrium collision trajectory, but provides access to the thermodynamic responses from which constrained trajectories and fluctuations are constructed.

Hypernuclei provide a complementary empirical probe of the response to strangeness. These are bound nuclear systems containing one or more hyperons, which are stable under the strong interaction and decay weakly. A recent analysis of experimental nuclear and hypernuclear binding energies ~\cite{Noronha-Hostler:2026ykk} extracted mesoscopic analogues of the baryon, charge, and strangeness chemical potentials \footnote{Formally, these are finite-difference slopes of the strong-interaction energy landscape of finite nuclei, not equilibrium grand-canonical chemical potentials.}, providing empirical response scales that can be compared with homogeneous-matter \eos\ calculations near saturation. The extracted analogue chemical potentials fall in the range $\hat\mu_S \in [-185, -165]$~MeV, with $\hat\mu_S \approx -177$~MeV for nearly isospin-symmetric systems. 
This identifies a physically motivated range of negative strangeness chemical potentials precisely in the density regime associated with nuclear saturation and the liquid branch of the nuclear liquid--gas transition.

The influence of strangeness on the liquid--gas transition has previously been investigated in relativistic and non-relativistic models of nucleon--hyperon matter~\cite{Yang:2003up,Wang:2004gya,Raduta:2014lja,Torres:2015oqa}. These studies showed that hyperons can participate in the instability and that the coexisting phases generally carry different strange fractions, leading to a fractionation of strangeness between the dilute and dense phases. In multicomponent matter, the unstable mode also need not be aligned with the total baryon density, and the strangeness density can become part of the order parameter. Here, we instead follow the coexistence structure continuously as a function of the externally specified $\mu_S$. At fixed $\mu_Q$, the coexistence surface lies in $(T,\mu_B,\mu_S)$, with a critical line along its boundary. This allows us to determine how the composition of the coexisting phases, the $BQS$ direction of the order parameter, and the critical locus change as the strange direction is explored. The mesoscopic values extracted from hypernuclei serve as a phenomenological benchmark rather than a direct thermodynamic constraint.

A possible experimental setting for this transition is provided by spectator hypermatter. In peripheral relativistic heavy-ion collisions, hyperons produced in the participant region can be captured by the excited nuclear spectator, producing hypernuclear matter at excitation energies and temperatures characteristic of multifragmentation~\cite{Botvina:2007pd,Botvina:2011jt}. Statistical multifragmentation models have connected the breakup of these systems with the liquid--gas transition of hypermatter, and multistrange spectators are also predicted, although at substantially smaller rates. Hypernuclear production with relativistic beams has already been demonstrated by the HypHI experiment through the observation of ${}^3_\Lambda{\rm H}$ and ${}^4_\Lambda{\rm H}$ in $^6{\rm Li}+{}^{12}{\rm C}$ collisions~\cite{Rappold:2013fic}. Further experimental programs continue to explore the production and structure of such systems~\cite{Saito:2016ulw,Suvarieva:2016sya,Kapishin:2021lxs}. Complementary opportunities may be provided by PUMA. Although its primary program probes the neutron and proton content of nuclear-density tails, it has identified hypernuclei and in-medium hyperon--nucleon interactions as part of its broader research program~\cite{PUMA:2022ngr,Obertelli:2622466}.

\section{Formalism}
\label{sec:formalism}

The thermodynamic state is specified by the temperature and  the  chemical potentials associated with the relevant charges conserved under the strong interaction. For these work, they are
\begin{equation}\label{eq:x_def}
\boldsymbol{x}=(T,\mu_B,\mu_Q,\mu_S).
\end{equation}

\subsection{The Chiral Mean Field model}
\label{sec:cmf}

The CMF model was developed through Refs.~\cite{Papazoglou:1997uw,Papazoglou:1998vr,Dexheimer:2008ax,Dexheimer:2009hi} from a nonlinear realization of ${\textrm{SU}(3)_L \times SU(3)_R}$ symmetry. The current work follows the open-source implementation of Ref.~\cite{Cruz-Camacho:2024odu}, together with the redefined vector fields of Ref.~\cite{Kumar:2024owe}, and extends the framework of Ref.~\cite{Cruz-Camacho:2024odu} to the finite-temperature thermodynamics described below. The baryon octet is represented by the flavor-space matrix $B$, while $q=(u,d,s)^T$ denotes the quark field vector; these fields interact through the scalar, vector, pseudoscalar, and axial-vector meson multiplets $S$, $V_\mu$, $Y$, and $A_\mu$, respectively.
In homogeneous, isotropic, and parity-even matter, the pseudoscalar and axial-vector fields have vanishing expectation values. The remaining interactions are mediated by the scalar and vector mean fields, and chiral symmetry is reflected in the structure of the mesonic self-interactions, in the dynamical generation of baryon masses, and through the symmetry-breaking terms in the scalar sector.

The nonet scalar and vector meson matrices are diagonal in flavor space,
\begin{equation}
S = {\rm diag} \left( \frac{\sigma +\delta}{\sqrt{2}}, \frac{\sigma - \delta}{\sqrt{2}}, \zeta\right),
\end{equation}
and the vector nonet matrix is
\begin{equation}
    V ={\rm diag}\left(\frac{\omega + \rho}{\sqrt{2}}, \frac{\omega - \rho}{\sqrt{2}}, \phi\right).
\end{equation}
Here, $\sigma$ and $\omega$ are the non-strange isoscalar fields, $\delta$ and $\rho$ are the isovector fields, and $\zeta$ and $\phi$ are the hidden-strangeness fields. We use $\phi$ for the strange vector field and $\Phi$ for the deconfinement field introduced below.\footnote{We denote the scalar nonet by $S$ and use the superscript $S$ for scalar-sector couplings, whereas Ref.~\cite{Cruz-Camacho:2024odu} uses $X$.} The interaction between the fermions and the scalar fields generates the effective mass of fermion species $i$,
\begin{equation}
m_i^\ast = g_{\sigma i}\sigma +g_{\zeta i}\zeta+g_{\delta i}\delta + \Delta m_i,
\label{eq:m_eff}
\end{equation}
while the vector fields shift the effective chemical potentials,
\begin{equation}
\mu_i^*=\mu_i-g_{ \omega i}\omega-g_{\rho i}\rho-g_{\phi i}\phi\,.
\label{eq:mu_eff}
\end{equation}
Here, $\Delta m_i$ is an explicit bare mass contribution, and $\mu_i$ is the fermion chemical potential given by \Cref{eq:mu_bare}. We collect all possible stationary mean fields in
\begin{equation}
\bvarphi=(\sigma,\zeta,\delta,\omega,\phi,\rho,\Phi),
\label{eq:mean_field_vector}
\end{equation}
where the deconfinement field $\Phi$ is introduced below and is omitted when the quark sector is disabled. For each mesonic field $\varphi\in\{\sigma,\zeta,\delta,\omega,\phi,\rho\}$, the coupling $g_{\varphi i}$ is fixed by the underlying chiral couplings $g_8^M$, $g_1^M$, and $\alpha_M$, with $M\in\{S,V\}$ denoting the scalar or vector sector, respectively\footnote{For details, see Appendix C.2 of~\cite{Cruz-Camacho:2024odu}.}. The isospin structure of the $\delta$ and $\rho$ interactions is included in the corresponding species-dependent couplings.

The mesonic contribution to the grand potential is written as
\begin{equation}
U_{\rm mes}
=
U_{\rm sc}
+U_{\rm vec}
+U_{\rm SB},
\label{eq:U_meson}
\end{equation}
where $U_{\rm sc}$ contains the scalar self-interactions and the terms associated with the chiral anomaly, $U_{\rm vec}$ contains the vector masses and self-interactions, and $U_{\rm SB}$ explicitly breaks chiral symmetry. We work in the frozen-dilaton approximation, in which the dilaton field is fixed at its vacuum expectation value. The complete expressions are given in Refs.~\cite{Papazoglou:1998vr,Cruz-Camacho:2024odu,Kumar:2024owe}.

We consider the vector self-interaction structures discussed in Ref.~\cite{Dexheimer:2015qha}. For each preset vector self-interaction structure, one may choose either the original vector fields, denoted by C1--C4, or the redefined fields of Ref.~\cite{Kumar:2024owe}, denoted by RC1--RC4. In this paper, we focus on RC2 and RC4. In the tilded basis used to define the field-redefined interactions, their quartic invariants are
\begin{subequations}
\label{eq:vector_invariants}
\begin{align}
\widetilde{\mathcal L}_{\mathrm{vec,RC2}}^{\mathrm{SI}}
 &=
 \widetilde g_4
 \left\{
 \frac{3}{2}
 \left[\operatorname{Tr}\!\left(\widetilde V^2\right)\right]^2
 -
 \operatorname{Tr}\!\left(\widetilde V^4\right)
 \right\},
 \label{eq:vector_invariant_RC2}
 \\
\widetilde{\mathcal L}_{\mathrm{vec,RC4}}^{\mathrm{SI}}
 &=
 \frac{\widetilde g_4}{4}
 \left[\operatorname{Tr}\!\left(\widetilde V\right)\right]^4 .
 \label{eq:vector_invariant_RC4}
\end{align}
\end{subequations}
Using
\begin{equation}
 \widetilde\omega=\sqrt{Z_\omega}\,\omega,\qquad
 \widetilde\rho=\sqrt{Z_\rho}\,\rho,\qquad
 \widetilde\phi=\sqrt{Z_\phi}\,\phi
\end{equation}
we define
\begin{equation}
 r_\rho\equiv\frac{Z_\rho}{Z_\omega}=1,
 \qquad
 r_\phi\equiv\frac{Z_\phi}{Z_\omega},
\end{equation}
and we redefine the coupling as $g_4=Z_\omega^2\widetilde g_4$. 

The corresponding mean-field self-interactions are
\begin{subequations}
\label{eq:vector_self_interactions}
\begin{align}
\mathcal L_{\mathrm{vec,RC2}}^{\mathrm{SI}}
 ={}g_4\bigg[  &\omega^4  +r_\rho^2\rho^4  +\frac{r_\phi^2}{2}\phi^4  \label{eq:vector_self_interaction_RC2}\\ 
 &+3r_\rho r_\phi\rho^2\phi^2
 +3r_\phi\omega^2\phi^2
 \bigg], \nonumber
 \\
\mathcal L_{\mathrm{vec,RC4}}^{\mathrm{SI}}
 ={}g_4\bigg[&
 \omega^4
 +2\sqrt{2}\,r_\phi^{1/2}\omega^3\phi
 +3r_\phi\omega^2\phi^2
  \label{eq:vector_self_interaction_RC4} \\
 &+\sqrt{2}\,r_\phi^{3/2}\omega\phi^3 +\frac{r_\phi^2}{4}\phi^4
 \bigg].\nonumber
\end{align}
\end{subequations}
Thus, the RC2 vector self-interaction contains only even powers of the strange vector field $\phi$ and includes a $\rho^2\phi^2$ interaction, whereas the RC4 self-interaction contains no $\rho$-dependent terms and includes odd mixed $\omega$--$\phi$ contributions. In particular, the term proportional to $\omega^3\phi$ directly couples the nonstrange and strange isoscalar vector fields at finite baryon density. Their contributions to the mesonic potential appearing in \Cref{eq:U_meson} are $U_{\mathrm{vec},\mathcal C}^{\mathrm{SI}}
=-\mathcal L_{\mathrm{vec},\mathcal C}^{\mathrm{SI}}$, with 
$\mathcal C\in\{\mathrm{RC2},\mathrm{RC4}\}$.
The accompanying \cmfpp{} implementation also supports general linear combinations of the chiral vector invariants introduced in Ref.~\cite{Legred:2026cre}. 
For the baryon octet, $\Delta m_i$ contains a common bare-mass contribution
$m_0$. The RC2 parametrization uses $m_0=0$, whereas RC4 uses
$m_0=150$~MeV.

To include deconfined quarks, we introduce a Polyakov-loop-inspired field $\Phi$~\cite{Dexheimer:2009hi,Ratti:2005jh,Roessner:2006xn}. It couples to baryons and quarks through
\begin{equation}
\mathcal{L}_{\Phi}
=
-g_{B\Phi}\Phi^2 \mathrm{Tr}\left(\bar B B\right)
-g_{q\Phi}(1-\Phi)\bar{q}q,
\end{equation}
which modifies the effective masses according to
\begin{equation}
m_i^\ast \rightarrow m_i^\ast + \begin{cases}
 g_{B\Phi}\Phi^2\quad  {\textrm{for baryons}}\\ 
 g_{q\Phi}\left(1-\Phi \right) \quad {\textrm{for quarks}} \,.  
\end{cases}
\end{equation}
At small $\Phi$, the quarks acquire a large additional mass, while at large $\Phi$ the baryons are suppressed.

We consider a potential of the form
\begin{align}
U_{\Phi}&=\left(a_0T^4+a_1\mu_B^4+a_2T^2\mu_B^2\right)\Phi^2\nonumber \\ &+a_3T_0^4\ln\left(1-6\Phi^2+8\Phi^3-3\Phi^4\right)\,.
\label{eq:Phi_potential}
\end{align}
The logarithmic term restricts the allowed domain of $\Phi$ near its upper boundary. In the numerical calculation, we retain only solutions in the physical interval $0\leq\Phi<1$. The explicit temperature dependence primarily controls the location of the critical point and the shape of the deconfinement curve, while the explicit $\mu_B$ dependence allows additional control over the low-temperature, high-density regime. The $(T, \mu_B)$ dependence also contributes directly to the thermodynamic derivatives. Although this quark sector is available in the general \cmfpp{} implementation, we do not consider it in the calculations presented in this work.

The treatment of decuplet baryons, described by the Rarita--Schwinger fields, differs from the baryon-octet formalism~\cite{Zschiesche:2000ew}. Nevertheless, their effective masses and chemical potentials follow the form of \Cref{eq:mu_eff} and \Cref{eq:m_eff}. The construction of the decuplet couplings used here is summarized in \Cref{app:decuplet}.

\subsection{Thermodynamics and particle contents}
The grand-potential density at fixed $(T,\mu_B,\mu_Q,\mu_S)$ is
\begin{align}
\Omega
&=
 U_{\rm mes}
+ U_\Phi(T,\mu_B,\Phi)
-\Omega_{\rm vac}
\nonumber\\
&-
\sum_i
\frac{\gamma_i T}{2\pi^2}
\int_0^\infty k^2\,dk\,
\bigg[
\ln\left(1+e^{-(E_i^\ast-\mu_i^\ast)/T}\right)
\nonumber\\
&
+
\ln\left(1+e^{-(E_i^\ast+\mu_i^\ast)/T}\right)
\bigg].
\label{eq:Omega_potential}
\end{align}
where
\begin{equation}
E_i^\ast=\sqrt{k^2+m_i^{\ast 2}},
\end{equation}
with $\gamma_i=(2J_i+1)N_{c,i}$ the degeneracy of species $i$, where $J_i$ is its spin and $N_{c,i}=1$ for color-singlet hadrons or $3$ for quarks. The constant $\Omega_{\rm vac}$ is such that the vacuum pressure vanishes. The term $U_\Phi$ and the field $\Phi$ are omitted in calculations without quarks.

The mean fields are determined by the stationarity conditions
\begin{equation}
\frac{\partial\Omega}{\partial\varphi_a}=0,
\qquad a=1,\ldots,N.
\label{eq:stationary_condition}
\end{equation}
These equations can have more than one solution at the same thermodynamic point. We retain all solutions found by the root solver and classify them using the susceptibility matrix and pressure (discussed in the following).

The particle and antiparticle distributions are
\begin{equation}
    f_{i} = \frac{1}{1+ \exp{\left[\left(E^\ast_i - \mu_i^\ast\right)/T\right] } },
\end{equation}
and 
\begin{equation}
    \bar f_{i} = \frac{1}{1+ \exp{\left[\left(E^\ast_i + \mu_i^\ast\right)/T\right] } }.
\end{equation}
The antiparticle expression is equivalently obtained by assigning $\mu_{\bar i}^\ast=-\mu_i^\ast$, which reverses the conserved charges and the vector mean-field contribution. The form above is the corresponding compact convention when particles and antiparticles are combined under the same species label $i$.
Thermodynamic quantities are then obtained by integrating these distributions over momentum space $k \in (0, \infty)$, weighted by the relevant kinematic factors for each thermodynamic quantity.  The net number density is
\begin{equation}
n_i = \frac{\gamma_i}{2\pi^2} \int_0^\infty dk\,k^2 \left(f_i-\bar f_i\right),
\label{eq:nv_i}
\end{equation}
and the scalar density is
\begin{equation}
n_i^s =
\frac{\gamma_i}{2\pi^2}
\int_0^\infty dk\,k^2
\frac{m_i^\ast}{E_i^\ast}
\left(f_i+\bar f_i\right).
\label{eq:ns_i}
\end{equation}
The kinetic contribution to the energy density is
\begin{equation}
\varepsilon_i
=
\frac{\gamma_i}{2\pi^2}
\int_0^\infty dk\,k^2E_i^\ast
\left(f_i+\bar f_i\right).
\label{eq:energy}
\end{equation}
The meson fields additionally contribute to the energy density through the fermion-meson interactions and the meson self-interactions, as in Eqs.~(60) and (61) of~\cite{Cruz-Camacho:2024odu}. The total pressure is
\begin{equation}
p=-\Omega,
\end{equation}
which already contains the scalar and vector mesonic contributions through $U_{\rm mes}$.

Because $U_\Phi$ depends explicitly on $T$ and $\mu_B$, the baryon and entropy densities are
\begin{equation}
n_B
=
\left.\frac{\partial p}{\partial\mu_B}\right|_{T,\mu_Q,\mu_S}
=
\sum_i B_i n_i-\frac{\partial U_\Phi}{\partial\mu_B},
\end{equation}
and
\begin{equation}
s
=
\left.\frac{\partial p}{\partial T}\right|_{\mu_B,\mu_Q,\mu_S}
=
\sum_i s_i-\frac{\partial U_\Phi}{\partial T}.
\end{equation}

The eight preset CMF parametrizations available in the accompanying release of \cmfpp{} in MUSES are summarized in \Cref{tab:saturation_properties}. They are organized into four choices of vector self-interactions, each evaluated with the original and redefined vector fields (see Eq.~(14) of ~\cite{Cruz-Camacho:2024odu}).
\begin{table}
\caption{Nuclear saturation properties of the eight preset CMF parametrizations. We follow the MUSES notation of Ref.~\cite{Jahan:2026hvs}: $n_{\mathrm{sat}}$ is the saturation density, $E/A$ the energy per baryon relative to the vacuum nucleon mass, $K_0$ the incompressibility, $E_{\mathrm{sym}}$ the symmetry energy, and $L_{\mathrm{sym}}$ its slope.}
\label{tab:saturation_properties}
\setlength{\tabcolsep}{7.5pt}
\begin{tabular}{llllll}
\toprule
Model\  & $n_{\mathrm{sat}}$  & $E/A$ & $K_0$  & $E_{\mathrm{sym}}$  & $L_{\mathrm{sym}}$ \\
 &  [fm$^{-3}$] & [MeV] &  [MeV] &  [MeV] &  [MeV]\\
\midrule
RC1 & 0.1513 & -16.421 & 279.5 & 29.3 & 80.6 \\
C1 & 0.1451 & -16.487 & 273.9 & 30.2 & 66.9 \\
RC2 & 0.1513 & -16.421 & 279.5 & 29.3 & 101.6 \\
C2 & 0.1451 & -16.487 & 274.0 & 30.2 & 89.2 \\
RC3 & 0.1513 & -16.421 & 279.5 & 29.2 & 92.9 \\
C3 & 0.1451 & -16.487 & 274.0 & 30.2 & 79.5 \\
RC4 & 0.1514 & -16.358 & 306.9 & 29.0 & 87.9 \\
C4 & 0.1502 & -16.653 & 299.0 & 29.5 & 87.0 \\
\bottomrule
\end{tabular}
\end{table}
The differences among these parametrizations become more pronounced above saturation, where the different vector self-interaction structures and their accompanying parameter refits control the density dependence of the repulsion and the feedback from the particle populations.

\subsection{Thermodynamic stability and phase coexistence}
\label{sec:susceptibilities}

For the ordering $\boldsymbol{x}$ introduced in \Cref{eq:x_def}, we define the conjugate densities componentwise as
\begin{equation}
y_a
\equiv
\left.\frac{\partial p}{\partial x_a}\right|_{\{x_c\}_{c\ne a}},
\qquad
\boldsymbol{y}=(s,n_B,n_Q,n_S).
\end{equation}
Thus,
\begin{equation}
dp
=
s\,dT+n_B\,d\mu_B+n_Q\,d\mu_Q+n_S\,d\mu_S
=
y_a\,dx_a,
\end{equation}
Thermodynamic stability requires the pressure to be locally convex,
\begin{equation}
d^2p
=
d\boldsymbol{x}^{T}\boldsymbol{\chi}\,d\boldsymbol{x}
\geq0,
\qquad
d\boldsymbol{x}=(dT,d\mu_B,d\mu_Q,d\mu_S)^T,
\end{equation}
where the susceptibility matrix is
\begin{equation}
\boldsymbol{\chi} =
    \begin{pmatrix}
        \ch{TT} & \ch{TB} & \ch{TQ} & \ch{TS} \\
        \ch{BT} & \ch{BB} & \ch{BQ} & \ch{BS} \\
        \ch{QT} & \ch{QB} & \ch{QQ} & \ch{QS} \\
        \ch{ST} & \ch{SB} & \ch{SQ} & \ch{SS}
    \end{pmatrix},
\end{equation}
with
\begin{equation}
\chi_{ab}
=
\left.
\frac{\partial^2p}{\partial x_a\partial x_b}
\right|_{\{x_c\}_{c\ne a,b}}.
\end{equation}
See Appendix F of \cite{Cruz-Camacho:2024odu} for details. A homogeneous solution is locally stable when $\boldsymbol{\chi}$ is positive semidefinite.

For phase separation, it is useful to define the spinodal, which sets the local direction in which a homogeneous branch loses stability, and the binodal, which sets the global density difference between the two phases and where the Maxwell construction is usually applied, connecting the different stable phases. To define these, we isolate the fixed-temperature response of the conserved-charge densities,
\begin{equation}
\left(\boldsymbol{\chi}_T\right)_{ab}
=
\left.
\frac{\partial n_a}{\partial\mu_b}
\right|_{T,\{\mu_c\}_{c\ne b}},
\qquad
a,b,c\in\{B,Q,S\},
\end{equation}
and define the curvature matrix of the free-energy density with respect to the conserved-charge densities at fixed temperature,
\begin{equation}
\boldsymbol{C}_T=\boldsymbol{\chi}_T^{-1},
\end{equation}
whenever the inverse exists. Its smallest eigenvalue defines the soft curvature of the matrix, and the corresponding normalized eigenvector defines the soft direction,
\begin{equation}
\boldsymbol C_T\widehat{e}_{\rm soft}
=
\lambda_{\rm soft}\widehat{e}_{\rm soft}.
\end{equation}
The spinodal boundary is defined as the point where the soft eigenvalue reaches zero from the locally stable side    
\begin{equation}
\lambda_{\min}(\boldsymbol C_T)=0,
\label{eq:spinodal_condition}
\end{equation}
where the corresponding eigenvalue of $\boldsymbol{\chi}_T$ diverges.
Upon continuation into the unstable region, $\lambda_{\rm soft}$ becomes negative. The associated eigenvector gives the local direction in density space along which the homogeneous solution becomes unstable~\cite{Avancini:2006zi,Torres:2015oqa}.

Below the critical temperature, the lower- and higher-density spinodals define two local soft directions, which we denote by $\widehat{e}_{\rm soft}^{(g)}$ and $\widehat{e}_{\rm soft}^{(\ell)}$, respectively. The superscripts refer to the gas- and liquid-side branches of the liquid--gas transition and are retained as lower- and higher-density labels for the strangeness-dominated transition. These directions do not in general coincide away from the critical point.

The binodal is instead a global coexistence condition. At a common thermodynamic point $\boldsymbol{x}$, two distinct locally stable stationary solutions $g$ and $\ell$ satisfying \Cref{eq:stationary_condition} lie on the binodal when
\begin{equation}
 p^{(g)}(\boldsymbol{x})
 =
 p^{(\ell)}(\boldsymbol{x}) =
\max_k p^{(k)}(\boldsymbol{x}),
 \qquad
 \bvarphi^{(g)}\ne\bvarphi^{(\ell)}.
\label{eq:binodal_condition}
\end{equation}
The global direction of phase separation is characterized by the density discontinuity between these coexisting phases,
\begin{equation}
\Delta\boldsymbol n = \boldsymbol n^{(\ell)}-\boldsymbol n^{(g)}
= (\Delta n_B,\Delta n_Q,\Delta n_S)^T,
\label{eq:density_discontinuity}
\end{equation}
where the phases are ordered such that $\Delta n_B>0$. Its normalized direction is
\begin{equation}
\widehat{\Delta\boldsymbol n} = 
    \frac{\Delta\boldsymbol n}{|\Delta\boldsymbol n|}.
\label{eq:phase_separation_direction}
\end{equation}
The vector $\widehat{\Delta\boldsymbol n}$ connects the two phases on the
binodal, whereas $\widehat{e}_{\rm soft}^{(g)}$ and $\widehat{e}_{\rm soft}^{(\ell)}$ describe the local loss of
stability at the two spinodal boundaries. Away from the critical point, these three directions need not be the same.

For a critical point governed by a single soft mode, the two coexisting phases and the two spinodal boundaries merge. Their directions consequently approach the same critical mode,
\begin{equation}
\left(\widehat{\Delta\boldsymbol n}, \quad \widehat{e}_{\rm soft}^{(g)}, \quad
\widehat{e}_{\rm soft}^{(\ell)} \right)
\longrightarrow 
\widehat{e}_{c} \qquad \text{as} \qquad T\rightarrow T_c^-,
\label{eq:critical_direction_limit}
\end{equation}
while $|\Delta\boldsymbol n|\rightarrow0$~\cite{Ducoin:2005aa,Avancini:2006zi,Torres:2015oqa}. The normalized density discontinuity is therefore understood at the critical point through this limiting direction.

Finally, the displacement of the coexistence surface follows from equality of the pressures. Differentiating $p_1=p_2$ gives
\begin{equation}
\Delta s\,dT
+
\Delta n_B\,d\mu_B
+
\Delta n_Q\,d\mu_Q
+
\Delta n_S\,d\mu_S
=0.
\label{eq:generalized_cc}
\end{equation}
Along the critical locus at fixed $\mu_Q$, the limiting form of this relation becomes
\begin{equation}
\frac{d\mu_{B,c}}{d\mu_S}
=
-\frac{\Delta n_S}{\Delta n_B}
-
\frac{\Delta s}{\Delta n_B}
\frac{dT_c}{d\mu_S}.
\label{eq:cc_decomposition}
\end{equation}
This separates the displacement of the critical baryon chemical potential into a composition contribution and a thermal contribution.

\subsection{Jacobian calculation of the susceptibilities}

Susceptibilities are often calculated by taking numerical derivatives of thermodynamic tables. This causes two problems in the present application. Across a first-order transition, numerical derivatives mix discontinuous branches and generate large values that can be mistaken for a physical divergence. In addition, a lower-dimensional table does not contain the information needed to reconstruct the complete susceptibility matrix. A scan in $\mu_B$, for example, gives derivatives with respect to $\mu_B$, but it does not determine the response to independent charge and strangeness fluctuations. We instead use the Jacobian method of Ref.~\cite{Gholami:2025cfq}. We define the field-equation residuals as
\begin{equation}\label{eq:field_residuals}
F_i(\boldsymbol{x},\bvarphi)
\equiv
\left.
\frac{\partial\Omega}{\partial\varphi_i}
\right|_{\boldsymbol{x},\{\varphi_j\}_{j\ne i}}.
\end{equation}
The stationary solutions satisfy $\boldsymbol{F}=0$. Along each such solution, the susceptibilities are
\begin{equation}
\chi_{ab}
=
\left.
\frac{\partial y_a}{\partial x_b}
\right|_{\boldsymbol{F}=0,\{x_c\}_{c\ne b}}
=
\frac{H_{ab}}{D},
\label{eq:chi_jac}
\end{equation}
where
\begin{equation}
H_{ab}
=
\frac{\partial(y_a,F_1,\ldots,F_N)}
{\partial(x_b,\varphi_1,\ldots,\varphi_N)},
\ \ \ 
D
=
\frac{\partial(F_1,\ldots,F_N)}
{\partial(\varphi_1,\ldots,\varphi_N)}.
\end{equation}
Here, $H_{ab}$ and $D$ denote Jacobian determinants. In $H_{ab}$, all thermodynamic variables other than $x_b$ are held fixed; in $D$, all thermodynamic variables are held fixed. The integer $N$ is the number of active mean fields in \Cref{eq:mean_field_vector}, while the dilaton field $\chi$ is fixed at its vacuum value. If $\Phi$ is fixed or the quark sector is disabled, it is omitted from these Jacobians.

For the second-order susceptibilities, \Cref{eq:chi_jac} can be written as a Schur complement. We define
\begin{equation}
\begin{aligned}
\mathcal{A}_{ab}
&= -\left.\frac{\partial^2\Omega}{\partial x_a\partial x_b}\right|_{\bvarphi},
&\mathcal{M}_{ai}
&= \frac{\partial^2\Omega}{\partial x_a\partial\varphi_i},\\
&&\mathcal{H}_{ij}
&= \frac{\partial^2\Omega}{\partial\varphi_i\partial\varphi_j}.
\end{aligned}
\label{eq:def_schur}
\end{equation}
In $\mathcal A$, the mean fields and thermodynamic variables not differentiated are held fixed. In $\mathcal M$, the remaining thermodynamic variables and mean fields are held fixed, while in $\mathcal H$ all thermodynamic variables and the remaining mean fields are held fixed. The derivatives are taken by treating $\boldsymbol{x}$ and $\bvarphi$ as independent variables and are then evaluated at the stationary solution. The explicit $T$ and $\mu_B$ dependence of $U_\Phi$ is included in the corresponding elements of $\mathcal A$ and $\mathcal M$.

Using the determinant identity for a block matrix gives
\begin{equation}
H_{ab} = D \left[ \mathcal{A}_{ab} + \mathcal{M}_{ai}
\left(\mathcal{H}^{-1}\right)_{ij} \mathcal{M}_{bj} \right].
\end{equation}
Therefore,
\begin{equation}
\chi_{ab} = \mathcal{A}_{ab} +
\mathcal{M}_{ai} \left(\mathcal{H}^{-1}\right)_{ij} \mathcal{M}_{bj}.
\label{eq:chi_schur}
\end{equation}
The first term is the response obtained at fixed mean fields. The second contains the implicit response of the stationary fields to changes in the thermodynamic variables. This form is numerically more convenient than the determinant ratio and is the one used in the implementation.

\Cref{eq:chi_schur} provides the complete susceptibility matrix at each stationary solution, independently of the dimension of the thermodynamic grid. Although the Jacobian construction can be extended to higher-order derivatives, the expressions become increasingly complicated for the number of fields in the CMF model. Since our purpose here is to determine the local stability of the different phases, we restrict the present calculation to second order.

\subsection{Numerical implementation and software release}
\label{sec:numerical}

At $T=0$, the particle densities, energy density, pressure, and the derivatives entering the susceptibilities are evaluated using their analytic expressions. At finite temperature, the momentum integrals are evaluated in \cmfpp{} with the Cubature library~\cite{cubature}. The semi-infinite interval $k\in[0,\infty)$ is mapped to a finite integration domain, and the scalar and vector densities are evaluated together in a vector-valued integral. The same method is used for the derivatives of the densities with respect to $m_i^\ast$, $\mu_i^\ast$, and $T$ that enter \Cref{eq:chi_schur}. For the calculations presented here, we use relative and absolute integration tolerances of $10^{-8}$, with no limit on the number of function evaluations.

The field equations are solved using a C++ implementation of the modified Powell hybrid method of MINPACK~\cite{BurkardtFSOLVE}. The Jacobian used by the nonlinear solver is calculated by forward differences. The solver terminates when the estimated relative error in the fields is smaller than $10^{-12}$. We additionally require every component of the field-equation residual $F_i$ defined in \Cref{eq:field_residuals} to satisfy
\begin{equation}
\max_i\left|F_i\right| <  10^{-4}.
\end{equation}
Solutions outside the physical field domain are discarded. In particular, we require $0\leq\Phi < 1$, $\sigma\leq 0$, and $\zeta\leq0$.

Several stationary solutions can exist at the same thermodynamic point. To find them, the field equations are solved independently from a fixed set of initial field configurations covering the vacuum, nuclear, strangeness-rich, and quark deconfined regions. The same set is used at every point of the thermodynamic grid. Initial conditions may either be generated from ranges specified in the configuration file or supplied through an external file. Solutions reached from different initial conditions are compared in mean-field space, and repeated roots are kept only once. All distinct solutions satisfying the domain and residual conditions are written to the output.

The stationary solutions are subsequently organized into branches along one-dimensional sections of the thermodynamic grid. Branches at neighboring points are matched by comparing their positions in the ($n_B,\Phi$) plane. A branch is allowed to be absent for up to 15 consecutive grid points before its phase identity is removed to account for occasional failures of the root solver to recover a particular solution from the available initial conditions. Along scans in increasing $\mu_B$, we also require the pressure and baryon density of the selected equilibrium branch not to decrease, consistently with $\chi_{BB}\geq0$ on a locally stable branch. A first-order transition is identified when the pressure-maximizing branch changes between two consecutive values of $\mu_B$. Since we do not solve explicitly for the equal-pressure point, the transition is bracketed as
\begin{equation}
\mu_B^{(j-1)} < \mu_B^{\rm tr} \leq \mu_B^{(j)},
\label{eq:mu_condition}
\end{equation}
where phase 1 is favored at $\mu_B^{(j-1)}$ and phase 2 is favored at $\mu_B^{(j)}$. The transition chemical potential stored by the code is $\mu_B^{(j)}$. Correspondingly, the densities reported on the two sides of the transition are evaluated at the lower and upper endpoints of this bracket, rather than at a common equal-pressure point. The uncertainty in the transition chemical potential is therefore set by the final grid spacing.

The critical point is determined through successive scans with decreasing temperature and chemical-potential spacings. At fixed ($\mu_S,\mu_Q$), the highest temperature at which a change of the stable branch is found provides the current estimate of the critical temperature. The temperature and $\mu_B$ intervals are then reduced around this point until the chosen grid resolutions are reached. We use a resolution of $\Delta T = \Delta \mu_B = 10^{-4}$~MeV.

The thermodynamic grid points and the initial-condition searches are evaluated independently and parallelized using OpenMP. The complete set of stationary solutions is first stored in a common table. Post-processing then produces separate tables for the stable, metastable, and unstable solutions, together with the detected transitions, particle-resolved properties, mean fields, and susceptibilities.

This work also accompanies the open-source release of \cmfpp{} \texttt{v2.0.0}, developed as part of the MUSES collaboration~\cite{ReinkePelicer:2025vuh,Jahan:2026hvs}. 

\section{Results}
\label{sec:results}

\begin{figure*}[t]
    \centering

    \includegraphics[width=1.0\linewidth]{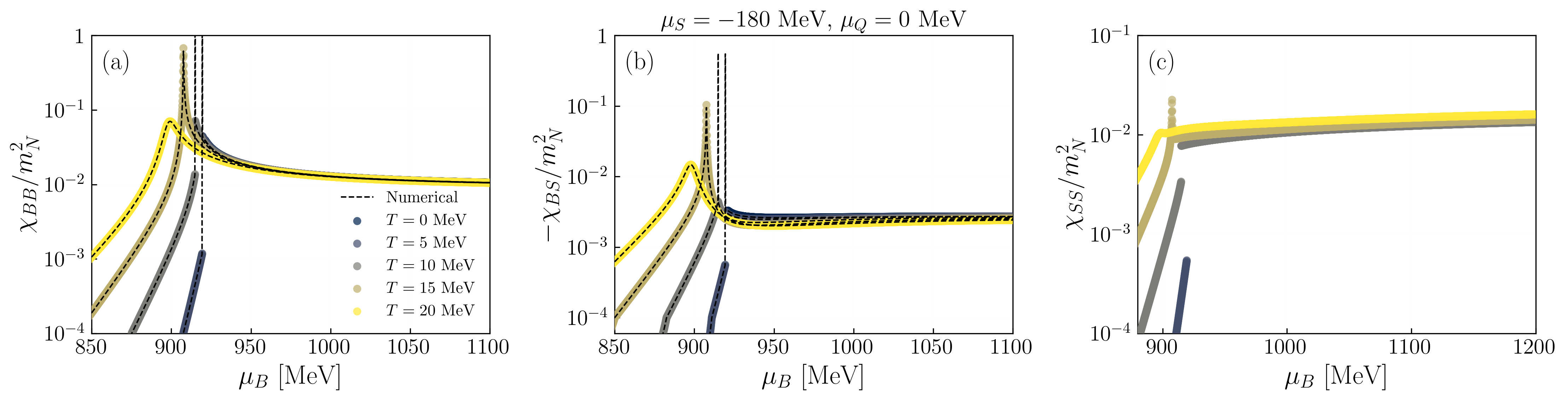}
    \caption{Second-order susceptibilities $\chi_{BB}$, $-\chi_{BS}$, and $\chi_{SS}$, divided by the nucleon mass squared, as a function of the baryon chemical potential in panels (a), (b), and (c), respectively. We fix $\mu_S=-180$ MeV and $\mu_Q=0$ and use the RC4 parametrization.  Colored curves show the Jacobian calculation. Dashed black lines show the finite-difference result in panels (a) and (b). The critical temperature is $14.46$ MeV.}
    \label{fig:LGY_chi}
\end{figure*}

In the following, we consider homogeneous, isospin-symmetric hadronic matter at fixed $\mu_Q=0$ and vary $(T,\mu_B,\mu_S)$. The quark sector and the deconfinement field $\Phi$ are disabled, while the baryon octet and decuplet are included throughout. We begin with the liquid--gas transition and follow its coexistence structure toward negative $\mu_S$ before turning to a separate high-density strangeness-dominated transition. We focus on the RC2 and RC4 parametrizations~\cite{Kumar:2024owe}. They share the same nuclear saturation point to good accuracy but differ in their vector self-interactions and in the population of strange and decuplet baryons, making them useful representatives of the model dependence.

The RC2 and RC4 parametrizations differ both in their quartic vector invariant and in the accompanying parameter fit. RC2 contains only even powers of the strange vector field $\phi$, whereas RC4 contains odd mixed terms, most notably $\omega^3\phi$, which directly couples the nonstrange and strange vector fields. Moreover, RC4 includes a finite common bare-mass contribution $m_0=150$~MeV to the baryon octet, while $m_0=0$ in RC2. The vector couplings and the explicit symmetry-breaking parameter $m_3^H$ are also refitted to describe the $\Lambda$ potential. Therefore, the different occurrence of the strangeness-dominated transition should be understood as sensitivity to the complete parametrization rather than attributed to a single coupling.

\subsection{The liquid--gas phase transition}
\label{res:liquid--gas}

In this section, we explore how the liquid--gas phase transition is affected by a finite strangeness chemical potential $\mu_S$. In Ref.~\cite{Noronha-Hostler:2026ykk}, a mesoscopic strangeness chemical-potential analog was extracted from hypernuclear binding energies, with values approximately in the range $\hat{\mu}_S\in[-185,-165]$~MeV. The experimental input used in that analysis consists of single- and double-$\Lambda$ hypernuclei. It therefore does not directly constrain baryons that themselves carry more than one unit of strangeness, such as the $\Xi$. Here, we extend the analysis to $\mu_S\in[-300,0]$~MeV to follow the reorganization of the transition as multistrange baryons become abundant. In particular, we study how the liquid-gas critical point $(T_c,\mu_{B,c})$ varies with $\mu_S$ and how the strange sector participates in the instability.
Our calculations do not include the surface, Coulomb, shell, or finite-size effects required for finite nuclei. Therefore, the hypernuclear values quoted above are phenomenological benchmarks. Extending this comparison to finite nuclei would require adding those effects explicitly. 

Before discussing the phase diagram, we validate the Jacobian-based susceptibility calculation and examine how the strange sector enters the liquid--gas response. \Cref{fig:LGY_chi} shows the baryon, mixed baryon--strangeness, and strangeness susceptibilities, divided by the nucleon mass squared, $m_N^2$, at $\mu_S=-180$~MeV and $\mu_Q=0$ for temperatures between $T=0$ and $20$~MeV. We choose $\mu_S=-180$~MeV to probe the regime suggested by the mesoscopic chemical-potential analogues. The finite-difference and Jacobian calculations generally agree along each continuous branch; their behavior at a discontinuous branch change is discussed below. Panel (c) shows only the Jacobian result because a finite-difference evaluation of $\chi_{SS}$ requires neighboring
points in $\mu_S$, whereas the numerical table used here is sampled in $(T,\mu_B)$ at fixed $\mu_S$. This illustrates one advantage of the Jacobian method.

Below the critical temperature, the equilibrium densities change discontinuously when the globally stable solution switches between the gas and liquid branches. A finite difference taken across this discontinuity mixes the two phases and produces a large, resolution-dependent value. This should not be interpreted as a physical divergence of the susceptibility. In contrast, the Jacobian method evaluates the local response of each stationary solution without differentiating across different branches. Along the continued branches, the susceptibility becomes singular at the spinodal boundary, while the physical critical divergence is recovered as the two coexisting phases and their spinodal boundaries merge at the critical point. The finite height of the calculated peaks reflects the finite resolution of the thermodynamic grid and the fact that the singular point is not sampled exactly. 

\begin{figure*}[t]
    \centering
    \includegraphics[width=0.4\linewidth]{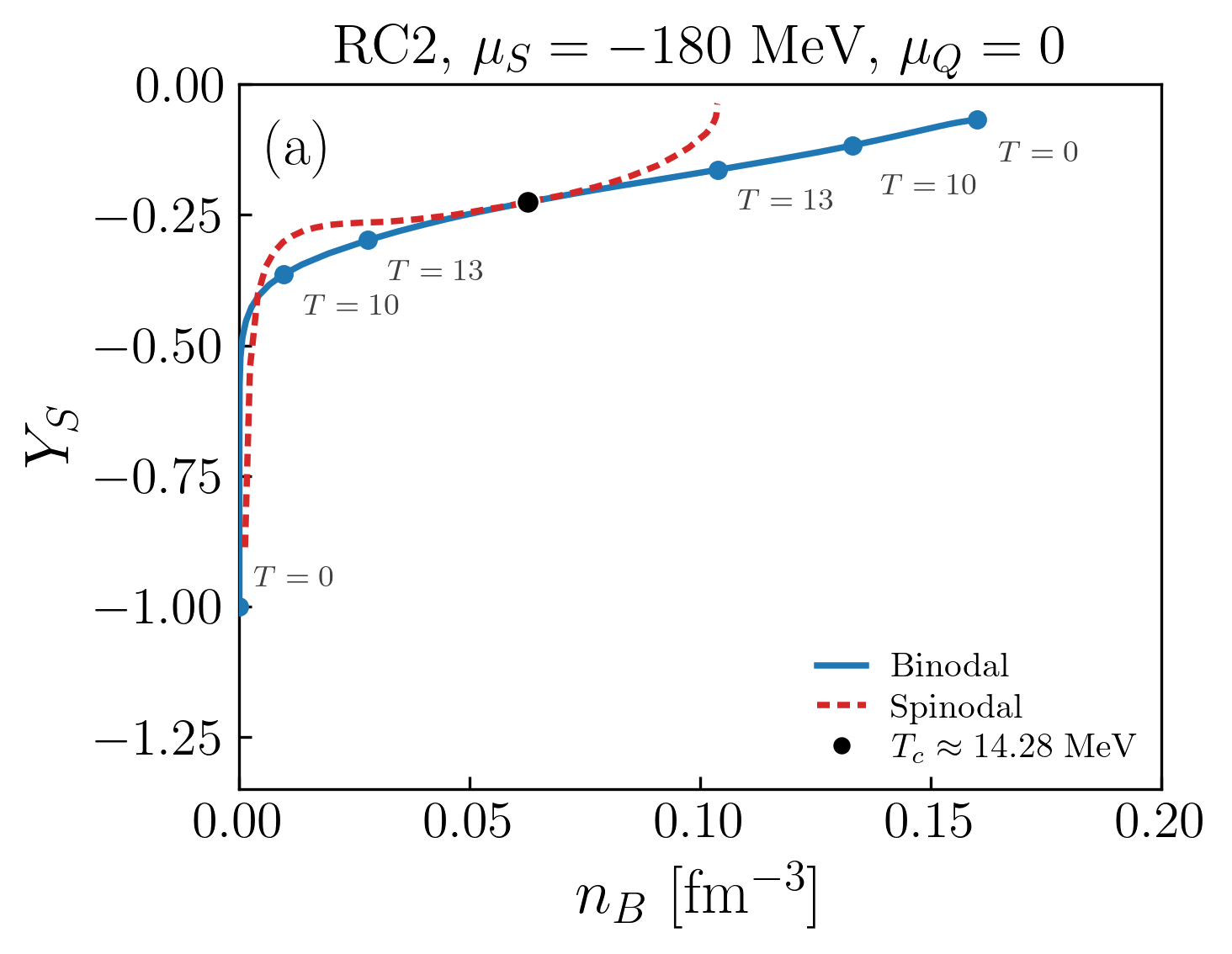}
    \includegraphics[width=0.4\linewidth]{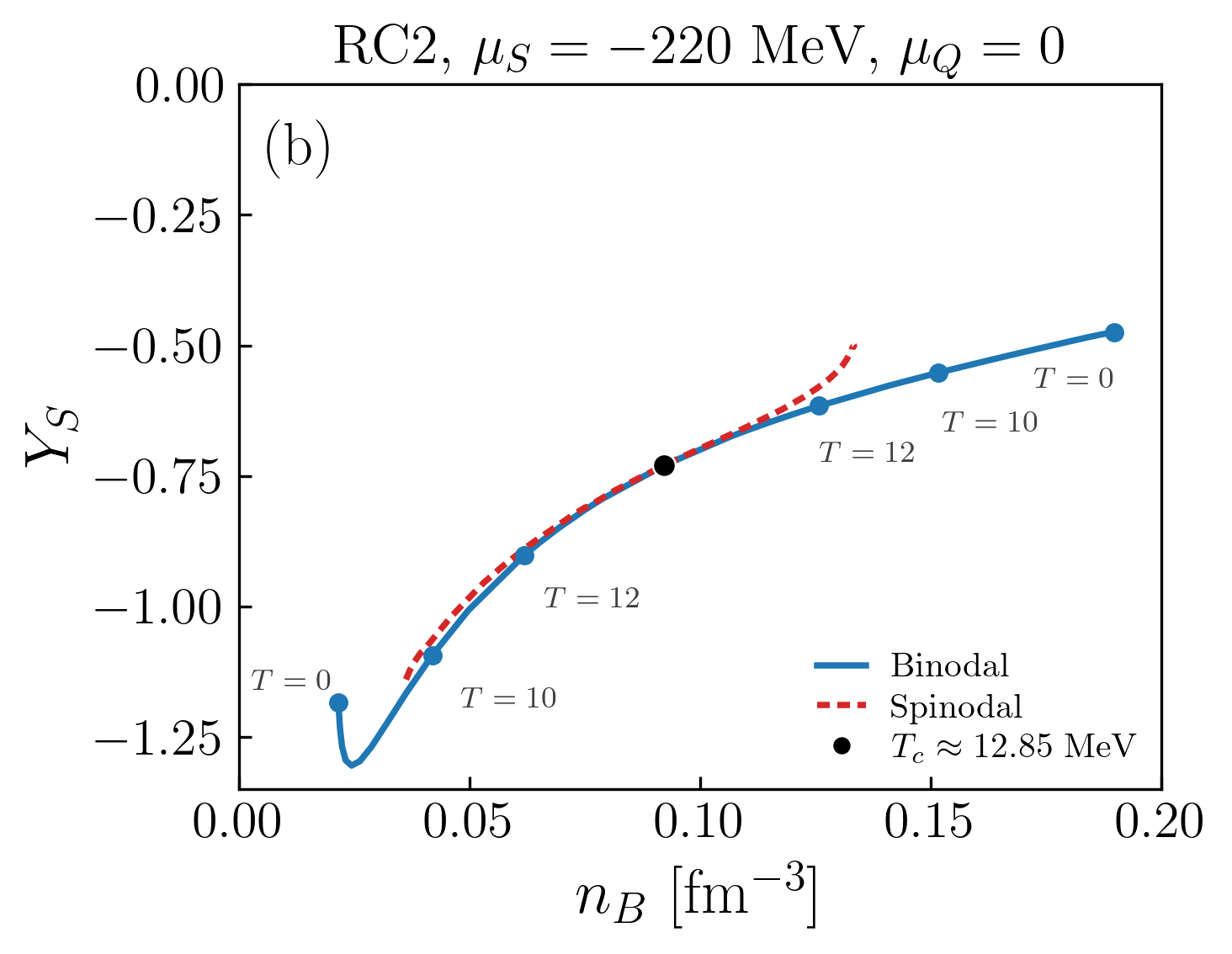}
    \includegraphics[width=0.4\linewidth]{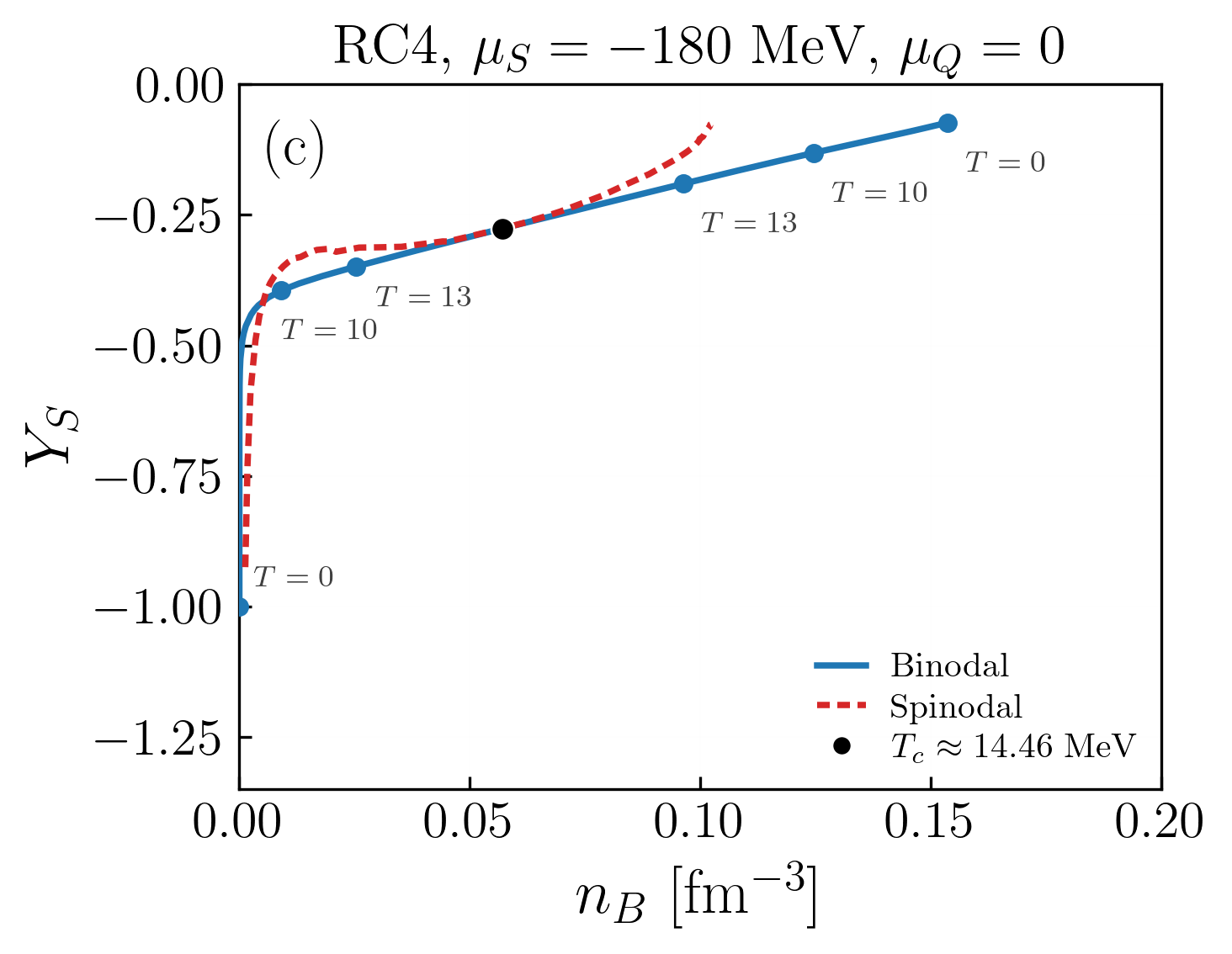}
    \includegraphics[width=0.4\linewidth]{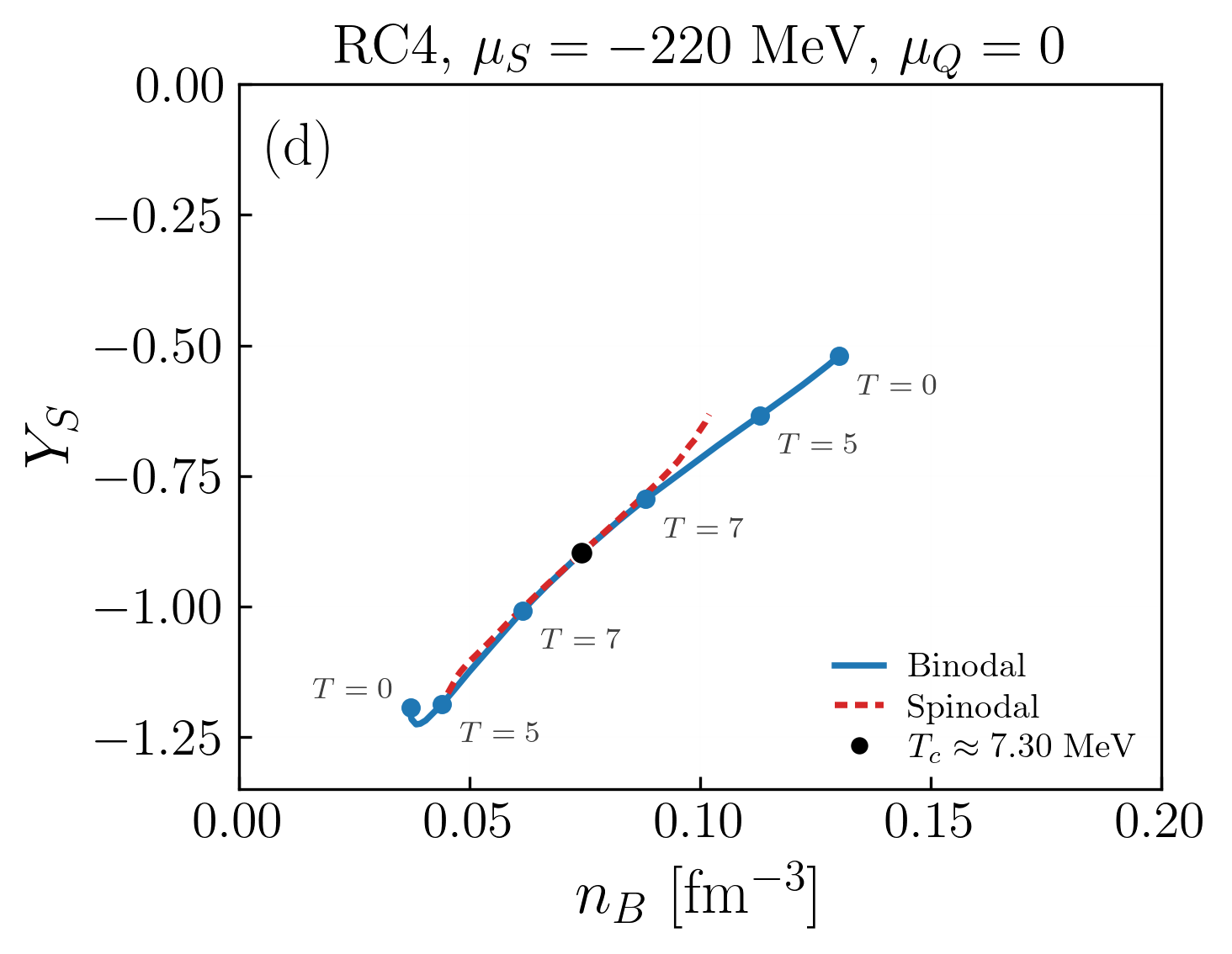}
\caption{Binodal (solid blue) and spinodal (dashed red) of the liquid--gas transition projected onto the $(n_B,Y_S)$ plane, with $Y_S=n_S/n_B$. Panels (a) and (b) show RC2 at $\mu_S=-180$ and $-220$~MeV, while panels (c) and (d) show the corresponding RC4 results. In all panels, $\mu_Q=0$. Blue circles identify selected temperatures along the binodal, and the black circle marks the critical point.}\label{fig:sp_bn_lgy}
\end{figure*}

\Cref{fig:LGY_chi} allows us to compare the influence of the liquid-gas phase transition on the different susceptibilities.
The mixed susceptibility $-\chi_{BS}$ follows the same qualitative structure as the baryonic response but remains smaller in magnitude. Its enhancement near the critical region shows that baryon and strangeness fluctuations are coupled. The corresponding structure in $\chi_{SS}$ is much weaker and is not resolved at the same scale, indicating that the critical response remains predominantly baryonic at $\mu_S=-180$~MeV.

Next, we examine the composition of the liquid-gas phase transition using its projection onto the $(n_B,Y_S)$ plane in \Cref{fig:sp_bn_lgy}. The binodal satisfies the equal-pressure coexistence condition in \Cref{eq:binodal_condition}, whereas each spinodal is the local-stability boundary defined by
\Cref{eq:spinodal_condition}.
The critical point is shown in black in each panel, together with the two coexistence branches for the corresponding parametrization and value of $\mu_S$.
The lower- and higher-density branches of the binodal correspond to the gas and liquid phases, respectively. Two points at the same temperature, one on each branch, form a coexistence pair. The spinodal lies inside the binodal and marks the boundaries of local stability. 

At $\mu_S=-180$~MeV, RC2 and RC4 give rather similar coexistence regions. In both cases, the dilute branch approaches $Y_S\simeq-1$ as $n_B\rightarrow0$ at low temperature, while the dense phase has a smaller strange fraction. The limit $Y_S\rightarrow-1$ means that the asymptotically small baryonic component carries one unit of negative strangeness per baryon, consistent with dominance by $S=-1$ hyperons ($\Lambda$'s).
Since the system is isospin symmetric at $\mu_Q=0$, the remaining charge fraction $Y_Q=n_Q/n_B$ follows from $Y_Q=(1+Y_S)/2$, so the same dilute limit has $Y_Q\rightarrow0$, whereas the denser, nearly nonstrange liquid approaches $Y_Q\simeq0.5$.

\begin{figure*}[t]
    \centering
    \includegraphics[width=0.95\linewidth]{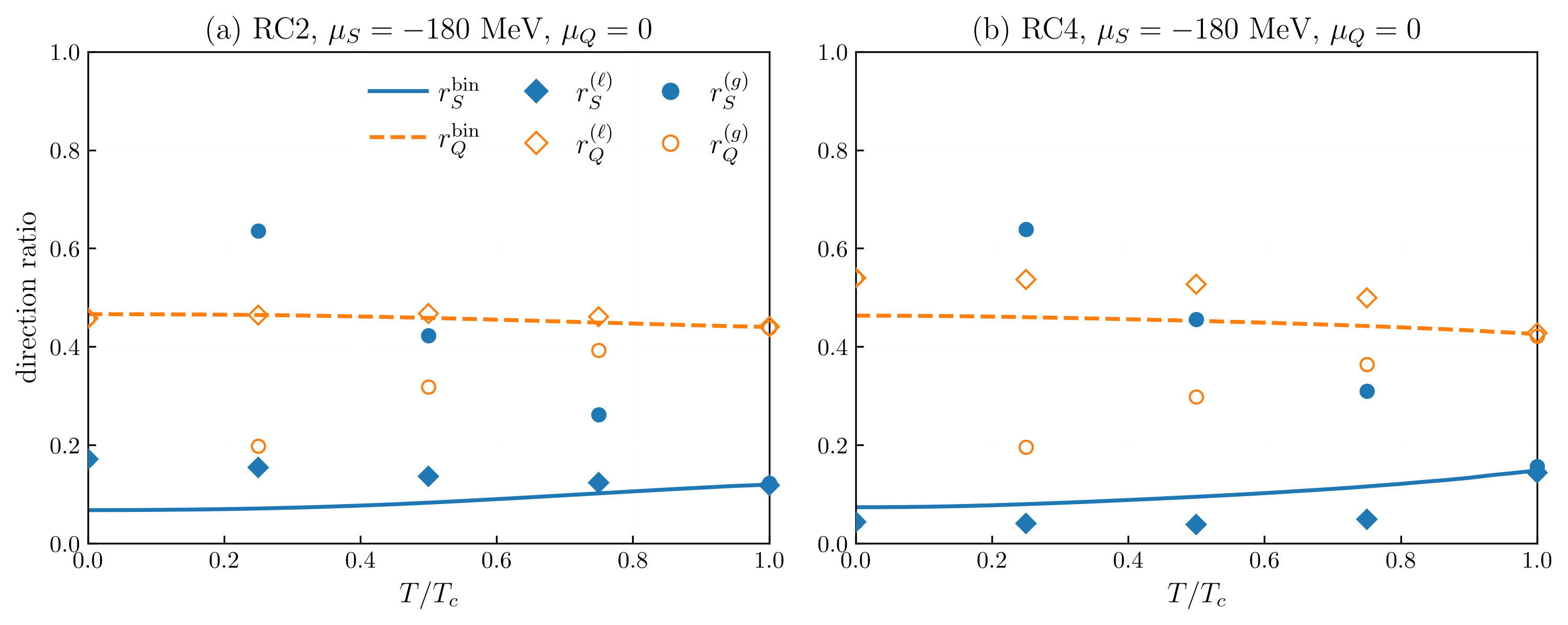}
    \includegraphics[width=0.95\linewidth]{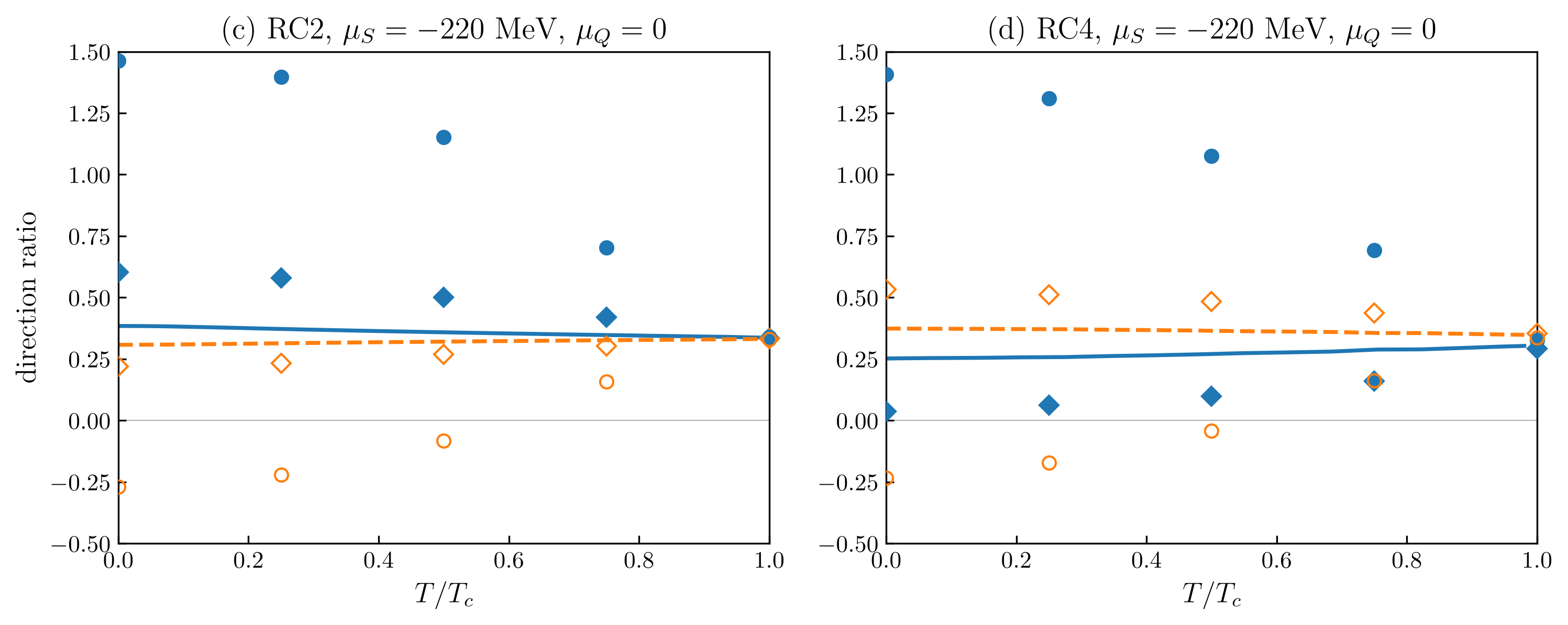}
    \caption{Baryon-normalized strangeness and charge components of the binodal density discontinuity and of the soft eigenvectors at the gas- and dense-side spinodals as functions of $T/T_c$. Panels (a) and (b) show RC2 and RC4 at $\mu_S=-180$~MeV, while panels (c) and (d) show the corresponding results at $\mu_S=-220$~MeV. The binodal direction is shown by the blue solid and orange dashed curves; circles and diamonds denote the gas- and dense-side spinodal directions, respectively. }    \label{fig:lgy_op}
\end{figure*}

The two parametrizations predict similar critical temperatures at $\mu_S=-180$~MeV, $T_c\simeq14.28$~MeV for RC2 and $T_c\simeq14.46$~MeV for RC4. At $\mu_S=0$, the corresponding values are approximately $14.9$ and $16.4$~MeV, respectively.

The difference between the parametrizations becomes much larger at $\mu_S=-220$~MeV, which demonstrates an even larger preference for strangeness. Both coexisting phases acquire substantially more strangeness, and the dilute branch reaches $Y_S<-1$. Such values cannot be produced by $S=-1$ baryons alone and correspond to a significant contribution from the $\Xi$ baryons. The critical point simultaneously moves toward a more strange and higher-density state. For RC2, the transition remains comparatively robust, with $T_c\simeq12.85$~MeV. In RC4, however, the coexistence region shrinks considerably and the critical temperature falls to $T_c\simeq7.30$~MeV. The behavior of the liquid--gas transition therefore becomes strongly sensitive to the parametrization once multistrange baryons contribute appreciably.

\begin{figure*}[t]
    \centering
    \includegraphics[width=0.48\linewidth]{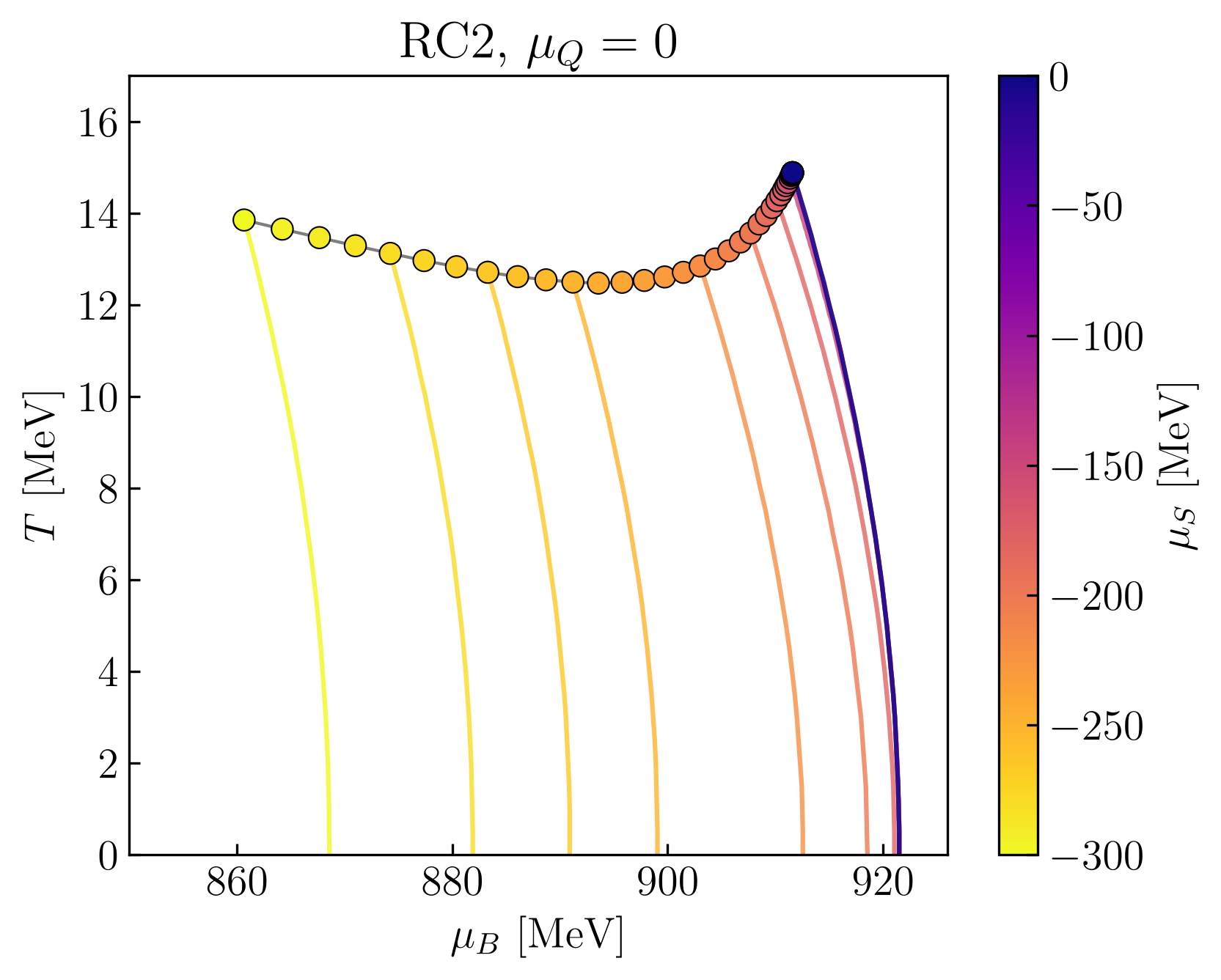}
    \includegraphics[width=0.48\linewidth]{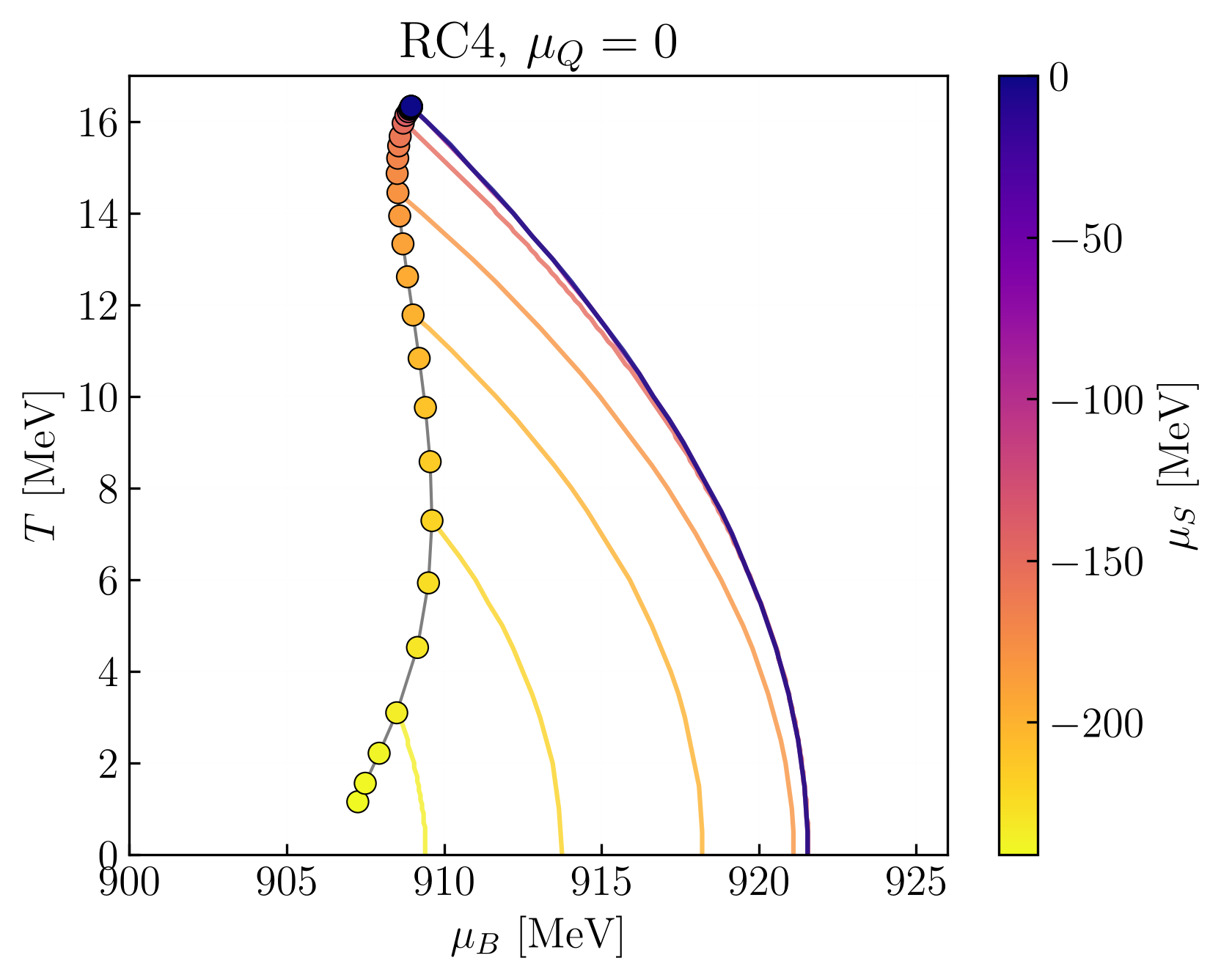}
\caption{Liquid--gas coexistence lines in the $(\mu_B,T)$ plane for RC2 (left) and RC4 (right) at $\mu_Q=0$. The color denotes $\mu_S$, and the circles mark the endpoints of the coexistence lines, thereby tracing the critical locus.}    \label{fig:critical_loci_LGY}
\end{figure*}

To identify which conserved-charge densities participate in the instability and compare the order-parameter direction directly with the composition term in the Clausius--Clapeyron slope, we introduce baryon-normalized direction ratios. We characterize the local spinodal directions through the soft eigenvectors $\widehat{e}_{\rm soft}^{(k)}\equiv(e_B^{(k)},e_Q^{(k)},e_S^{(k)})$ and the direction of phase coexistence through the density discontinuity $\Delta\boldsymbol n$, \Cref{eq:phase_separation_direction}. Their charge and strangeness components are

\begin{equation}
r_Q^{\rm bin}
=
\frac{\Delta n_Q}{\Delta n_B},
\qquad
r_S^{\rm bin}
=
-\frac{\Delta n_S}{\Delta n_B},
\label{eq:binodal_direction_ratios}
\end{equation}
and
\begin{equation}
r_Q^{(k)} = \frac{e_Q^{(k)}}{e_B^{(k)}},
\qquad
r_S^{(k)} = -\frac{e_S^{(k)}}{e_B^{(k)}},
    \qquad k\in\{g,\ell\},
\label{eq:spinodal_direction_ratios}
\end{equation}
where the signs of the spinodal eigenvectors are fixed by requiring $e_B^{(k)}>0$. 
At $\mu_Q=0$, the charge and strangeness ratios are not independent and obey
\begin{equation}
r_Q=\frac{1-r_S}{2}.
\label{eq:isoscalar_direction}
\end{equation}
The minus sign follows directly from the isoscalar condition. Using $Q=I_3+(B+S)/2$ and $\Delta n_{I_3}=0$ for the third component of isospin gives $\Delta n_Q=(\Delta n_B+\Delta n_S)/2$; with $r_S=-\Delta n_S/\Delta n_B$, this becomes $r_Q=(1-r_S)/2$. The same derivation applies to the components of the soft eigenvectors.
We nevertheless show both ratios in \Cref{fig:lgy_op}, since their agreement with  \Cref{eq:isoscalar_direction} demonstrates directly that the phase-separation and soft-mode directions remain isoscalar throughout the coexistence region.

Away from the critical point, the binodal and spinodal $BQS$ directions can be quite different. The density discontinuity varies relatively smoothly with temperature, whereas the two local soft modes undergo a much stronger rotation. This is particularly clear on the gas-side spinodal at $\mu_S=-220$~MeV, where $r_S^{(g)}$ can exceed unity and $r_Q^{(g)}$ consequently becomes negative. The soft fluctuation then carries more than one unit of negative strangeness per unit baryon component and has a negative electric-charge component. This behavior is consistent with strong participation from the $S=-2$ sector, in particular $\Xi$ baryons.

As the critical point is approached, the gas- and liquid-side eigenvectors converge toward the direction of the binodal discontinuity. This provides a direct numerical realization of \Cref{eq:critical_direction_limit} and shows that the critical point is governed by a single soft mode. At $\mu_S=-180$~MeV, the strange component of this mode remains modest, with $r_S$ of order 0.1--0.15. At $\mu_S=-220$~MeV it increases to approximately 0.3, depending on the parametrization. Thus, although the critical direction remains dominated by its baryonic component, it acquires a substantial strange contribution as $\mu_S$ becomes more negative.

To interpret the appearance of strange baryons, consider first the zero-temperature threshold condition $\mu_i^*=m_i^*$. At finite temperature, particles can already have a nonzero thermal population below this crossing, so $\mu_i^*=m_i^*$ therefore marks the opening of a degenerate Fermi sea rather than the first appearance of the species. The following vacuum estimates consequently provide only orientation. For a $\Lambda$ baryon at $\mu_Q=0$, $\mu_\Lambda=\mu_B-\mu_S$ and, in the dilute limit where the vector mean fields vanish, $\mu_\Lambda^*\rightarrow\mu_B-\mu_S$. Neglecting in-medium shifts therefore gives
\begin{equation}
\mu_{B,\Lambda}^{\rm th}(\mu_S)\simeq m_\Lambda+\mu_S,
\end{equation}
whereas the corresponding neutron estimate is
\begin{equation}
\mu_{B,n}^{\rm th}\simeq m_n.
\end{equation}
Using the vacuum masses gives $\mu_{B,n}^{\rm th}\simeq939$~MeV and $\mu_{B,\Lambda}^{\rm th}\simeq936$ and $896$~MeV at $\mu_S=-180$ and $-220$~MeV, respectively. Thus, at $\mu_S=0$ the neutron is favored well before the $\Lambda$, while a sufficiently negative $\mu_S$ can make the $\Lambda$ favored over the neutron already in dilute matter. In the full CMF calculation, scalar attraction changes $m_i^*$ and vector interactions shift $\mu_i^*$, so these numerical estimates are not strict cutoffs and differ between RC2 and RC4.

\Cref{fig:critical_loci_LGY} shows how the critical point of the liquid--gas transition moves in the $(T,\mu_B)$ plane as $\mu_S$ is varied, together with selected coexistence lines. For $\mu_S\gtrsim-150$~MeV, the critical point changes only weakly because strange baryons are suppressed and the liquid--gas transition remains nearly nucleonic. At more negative $\mu_S$, strange and multistrange baryons become thermodynamically competitive and the RC2 and RC4 trajectories separate. In RC2, $\mu_{B,c}$ decreases steadily while $T_c$ passes through a shallow minimum near $\mu_S\simeq-245$~MeV. The particle populations show that this minimum occurs near the crossover $\mu_\Xi^*\simeq m_\Xi^*$, which at $T=0$ marks the opening of the $\Xi$ Fermi sea and at finite temperature marks the onset of a degenerate $\Xi$ population. The $\Xi$ is already thermally populated at less negative $\mu_S$, so this crossing does not mark its first appearance. Beyond it, the $\Xi$ fraction grows from about $0.4$ to $0.6$, the critical composition becomes $\Xi$ dominated, and $T_c$ rises again. In RC4, $T_c$ instead decreases rapidly as $\mu_S$ becomes more negative, while $\mu_{B,c}$ remains nearly stationary and develops a weak back-bending.

The different trajectories reflect the feedback of the changing composition on the self-consistent scalar and vector mean fields. In RC2, the nucleon effective mass at the critical point decreases from about $m_N^\ast \approx 775$ to $m_N^\ast \approx657$~MeV between $\mu_S=0$ and $-300$~MeV, indicating a strong scalar response. Over the same interval, the critical density increases from $0.055$ to $0.187$~fm$^{-3}$, $\mu_{B,c}$ decreases by about $50$~MeV, and the $\Xi$ fraction reaches approximately $0.6$. The transition therefore survives but turns into a denser, $\Xi$-dominated state. In RC4, by contrast, the nucleon effective mass remains near $800$~MeV and the critical density increases only from $0.057$ to $0.090$~fm$^{-3}$ between $\mu_S=0$ and $-240$~MeV. Strangeness is carried mainly by $\Lambda$ baryons over much of this interval, with the $\Xi$ fraction becoming comparable only near the end, while $T_c$ falls from $16.3$ to $1.6$~MeV and $\mu_{B,c}$ stays within a range of about $2.5$~MeV. Thus, the two parametrizations differ not only in when strange species become important, but also in how strongly their populations feed back on the scalar fields and sustain the liquid--gas transition.

\begin{figure*}[t]
    \centering
    \includegraphics[width=0.9\linewidth]{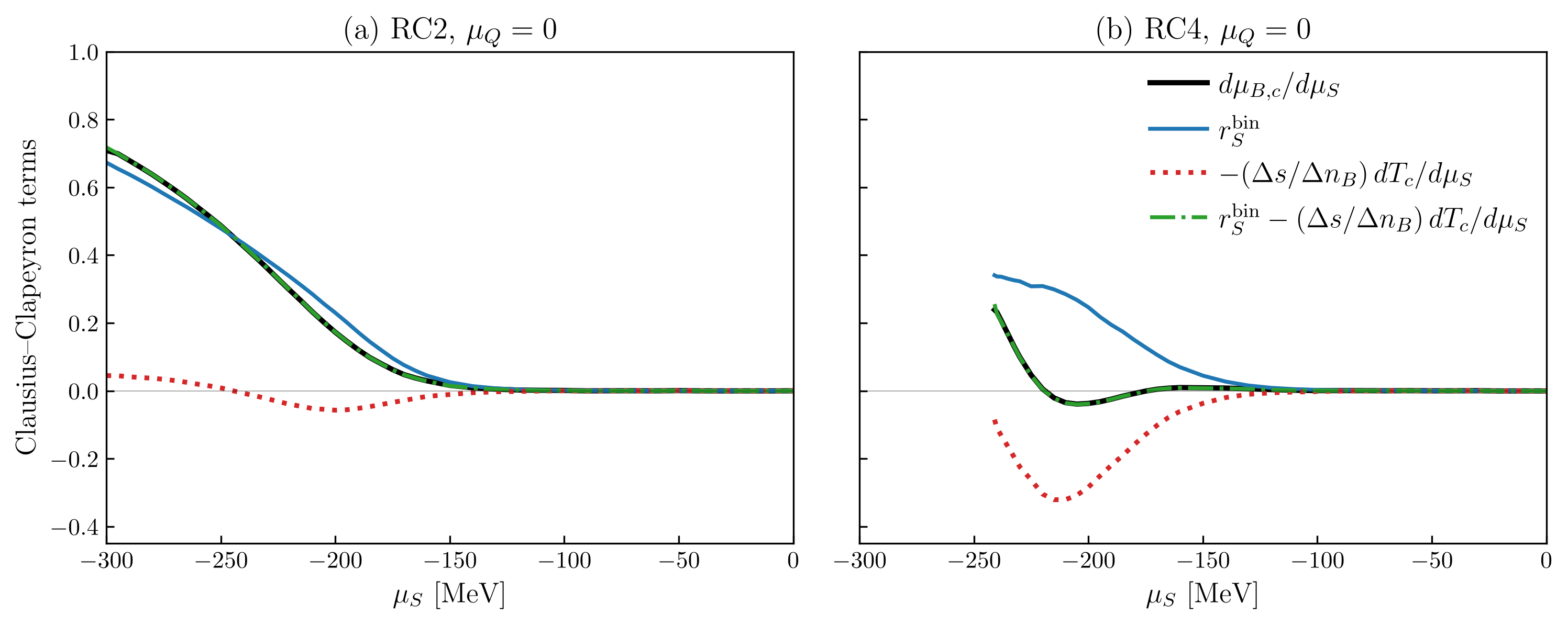}
    \caption{Generalized Clausius--Clapeyron decomposition of the liquid--gas critical locus for RC2 (left) and RC4 (right). The black curve shows $d\mu_{B,c}/d\mu_S$, while the blue solid and red dotted curves show the composition and thermal terms in \Cref{eq:cc_decomposition}. Their sum is shown by the green dash-dotted curve.}   
    \label{fig:lgy_cc}
\end{figure*}

The distinct RC2 and RC4 critical trajectories can be resolved with the generalized Clausius--Clapeyron relation, \Cref{eq:cc_decomposition}. \Cref{fig:lgy_cc} compares the directly evaluated slope $d\mu_{B,c}/d\mu_S$ with its composition contribution, $-\Delta n_S/\Delta n_B$, and its thermal contribution, $-(\Delta s/\Delta n_B)(dT_c/d\mu_S)$. The agreement between the direct derivative and the sum of the two terms provides an independent consistency check of the coexistence construction.

With the phase ordering $\Delta n_B>0$, the denser phase carries the more negative strangeness density, so $\Delta n_S<0$ and the composition term $-\Delta n_S/\Delta n_B$ is positive. The sign of the thermal term is instead set by the product $(\Delta s/\Delta n_B)(dT_c/d\mu_S)$. For $\mu_S\gtrsim-150$~MeV, strange particles are suppressed, $\Delta n_S$ is small, and $T_c$ varies weakly. Consequently, both contributions and the total slope are close to zero.

For RC2, the composition term controls the displacement of $\mu_{B,c}$. The small thermal term changes sign near the shallow minimum of $T_c$ but produces only a modest correction, so the trajectory is close to an isothermal composition-driven displacement. For RC4, the weak back-bending occurs between the $\Lambda$ and $\Xi$ Fermi-sea openings: $\mu_{B,c}$ has a shallow minimum near the $\Lambda$ crossing and reaches a shallow maximum near the $\Xi$ crossing. In this interval, the steep decrease of $T_c$ makes the thermal term large and negative around $\mu_S\simeq-220$~MeV, allowing it to cancel much of the positive composition term and briefly reverse the total slope. This separation between the blue composition curve and the black and green total-slope curves explains why the RC4 critical locus moves strongly in temperature while changing comparatively little in $\mu_B$. At the most negative end of the RC4 locus, the curves terminate because no further finite-temperature liquid--gas critical point is available from which these derivatives can be evaluated.

Taken together, these results show that negative $\mu_S$ continuously transforms the ordinary nuclear liquid--gas transition into a strangeness-coupled instability. The baryon-density component remains dominant at the critical point, but the participation of strangeness grows markedly once the multistrange sector becomes populated. The magnitude and thermodynamic consequences of this reorganization are strongly parametrization dependent.

\subsection{Strangeness-dominated transition}

We now turn to the strangeness-dominated transition found at high baryon density with the RC2 parametrization. This is the finite-temperature continuation of the high-density hadronic first-order transition found at $T=0$ in Ref.~\cite{Cruz-Camacho:2024odu}. At $\mu_Q=\mu_S=0$, that transition appeared for the C2 and C3 parametrizations and was absent for C1 and C4. Its trajectory through $(\mu_B,\mu_Q,\mu_S)$ space was subsequently investigated using C3. Here, we determine how the corresponding coexistence surface and critical boundary extend through $(T,\mu_B,\mu_S)$ space with RC2 at fixed $\mu_Q=0$. Here, we determine how its coexistence surface and critical boundary extend through $(T,\mu_B,\mu_S)$ space at fixed $\mu_Q=0$. We do not find a corresponding transition with RC4 over the range explored here, showing that its existence is even more sensitive than the liquid--gas transition to the model parametrization. Because the quark sector and the deconfinement field $\Phi$ are disabled, both coexisting phases are hadronic and the transition is distinct from deconfinement.

\begin{figure*}[t]
    \centering

    \includegraphics[width=1.0\linewidth]{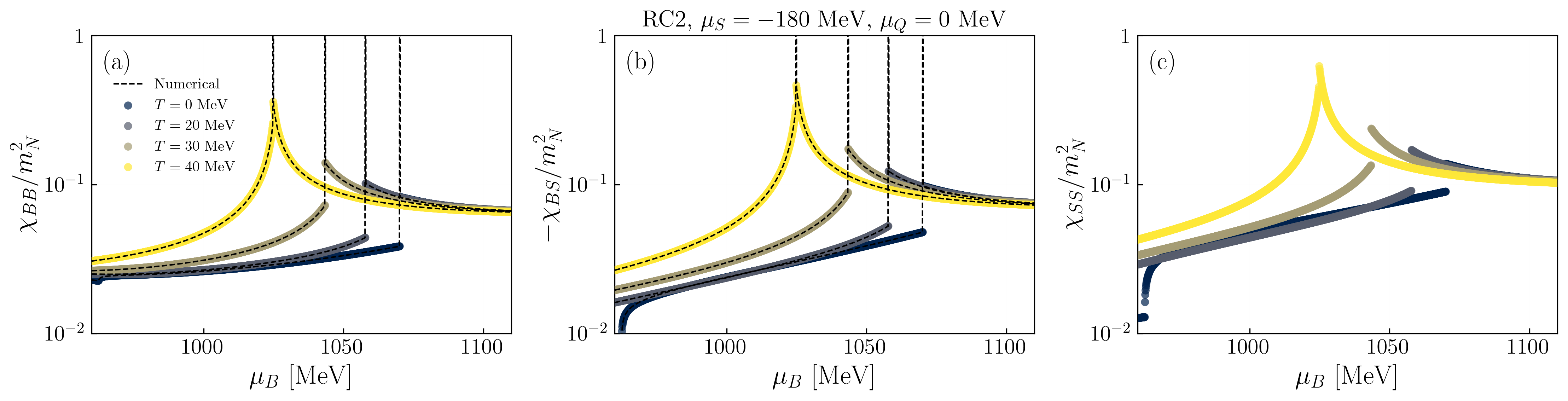}
    \caption{Second-order susceptibilities $\chi_{BB}$, $-\chi_{BS}$, and $\chi_{SS}$, divided by the nucleon mass squared, as functions of the baryon chemical potential in panels (a), (b), and (c), respectively. We use the RC2 parametrization at $\mu_S=-180$~MeV and $\mu_Q=0$. Colored curves show the Jacobian calculation, while dashed black lines show the finite-difference result in panels (a) and (b). The critical temperature is $45.2$ MeV.}
    \label{fig:STR_chi}
\end{figure*}

\Cref{fig:STR_chi} shows the susceptibility matrix elements at $\mu_S=-180$~MeV for temperatures between $T=0$ and $40$~MeV. As for the liquid--gas transition, the Jacobian and finite-difference calculations agree along each continuous branch, while a finite difference taken across the discontinuous change of branch produces a resolution-dependent spike. Over the transition region, the stable branches have $-\chi_{BS}/\chi_{BB}\simeq1.1$--$1.3$ and $\chi_{SS}/\chi_{BB}\simeq1.6$--$2.3$, with $-\chi_{BS}/\sqrt{\chi_{BB}\chi_{SS}}\simeq0.77$--$0.98$. Thus, $\chi_{SS}>-\chi_{BS}>\chi_{BB}$ and baryon-number and strangeness fluctuations are strongly anticorrelated, while the susceptibility matrix remains positive definite on each stable branch.

The discontinuity in a susceptibility at the phase transition measures the reorganization of the local linear response. At $T=0$, the branch-to-branch change is larger in $\chi_{BB}$ than in $\chi_{SS}$, indicating a larger contrast in baryonic compressibility, while the absolute $\chi_{SS}$ is larger on the lower-density branch. The order-parameter direction instead follows from the full susceptibility matrix near criticality and from the density discontinuity away from it, which we examine in the following.

\begin{figure*}[t]
    \centering
    \includegraphics[width=0.45\linewidth]{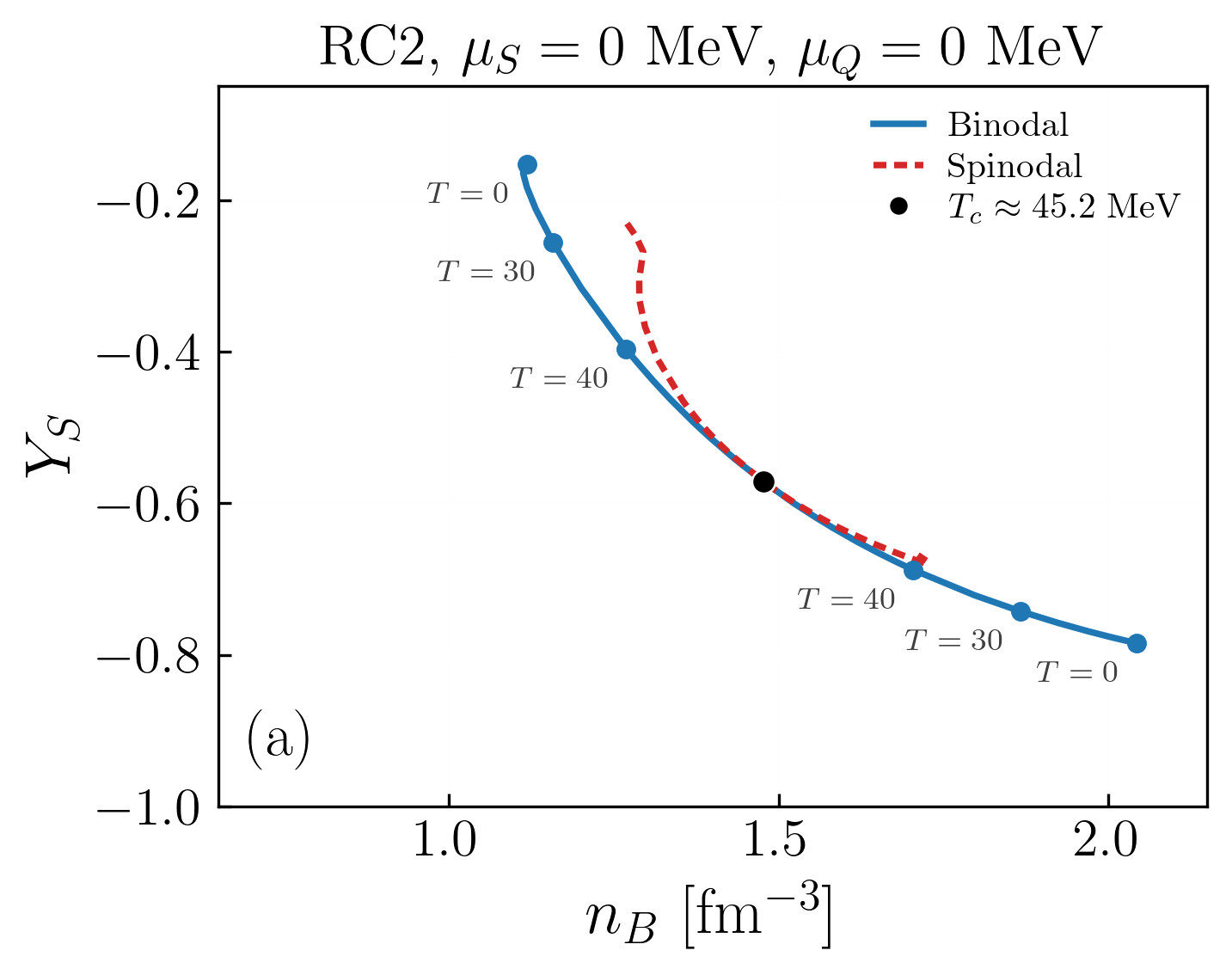}
    \includegraphics[width=0.45\linewidth]{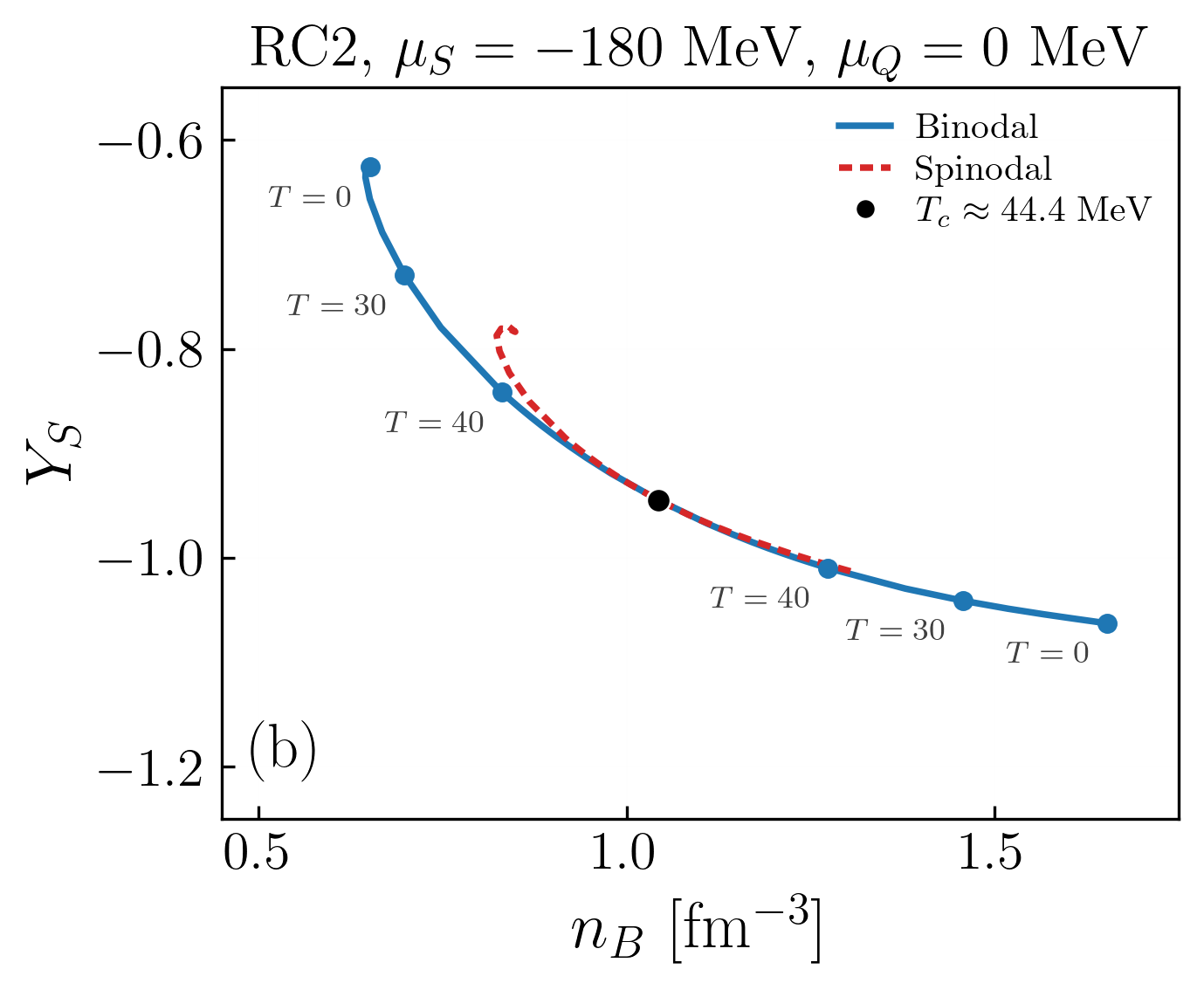}
    \caption{Binodal (solid blue) and spinodal (dashed red) of the strangeness-dominated transition projected onto the $(n_B,Y_S)$ plane at $\mu_Q=0$. Panels (a) and (b) show results at $\mu_S=0$ and $-180$~MeV, respectively. Blue circles identify selected temperatures along the binodal, and the black circle marks the critical point.}
    \label{fig:sp_bn_str}
\end{figure*}

We next examine the coexistence and spinodal boundaries in the $(n_B,Y_S)$ plane. \Cref{fig:sp_bn_str} shows the transition at $\mu_S=0$ and $-180$~MeV. In both cases, the two coexisting phases have markedly different strange compositions. Lowering $\mu_S$ favors strange baryons in both phases and moves the transition to smaller baryon densities. At $\mu_S=-180$~MeV, the dense phase reaches $Y_S<-1$, reflecting a substantial contribution from multistrange baryons.

At $\mu_Q=0$, isospin symmetry imposes $Y_Q=(1+Y_S)/2$. It therefore accommodates not only symmetric nucleonic matter, for which $(Y_S,Y_Q)=(0,1/2)$, but also increasingly strange compositions. Pure $\Lambda$ matter has $(Y_S,Y_Q)=(-1,0)$, while equal populations of $\Xi^0$ and $\Xi^-$ give $(Y_S,Y_Q)=(-2,-1/2)$. Consequently, a negative charge component of the density discontinuity does not indicate isospin breaking; it follows from the large negative-strangeness component of an isoscalar rearrangement.

The particle compositions show that, at $\mu_S=-180$~MeV and $T=0$, the lower-density phase is octet dominated, with fractions $Y_N\simeq0.60$, $Y_\Lambda\simeq0.18$, and $Y_\Xi\simeq0.23$. In the higher-density phase, decuplet baryons carry approximately $65\%$ of the baryon density: $Y_\Delta\simeq0.36$, $Y_{\Xi^*}\simeq0.22$, and $Y_\Omega\simeq0.07$. The $\Delta(1232)$ baryons supply about $59\%$ of $\Delta n_B$, whereas the contributions to $r_S^{\rm bin}=-\Delta n_S/\Delta n_B$ are approximately $0.73$ from $\Xi^*(1530)$, $0.36$ from $\Xi$, $0.35$ from $\Omega$, and $-0.10$ from $\Lambda$. Thus, $\Xi^*(1530)$ provides the largest contribution to the strangeness discontinuity. Across this transition, the $\Delta$, $\Xi^*$, and $\Omega$ Fermi seas open, whereas the $\Xi$ Fermi sea is already open on the lower-density branch. The compositions at $\mu_S=0$ and at the analyzed finite temperatures show the same qualitative reorganization from octet-dominated to decuplet-dominated matter.

\begin{figure*}[t]
    \centering
    \includegraphics[width=0.45\linewidth]{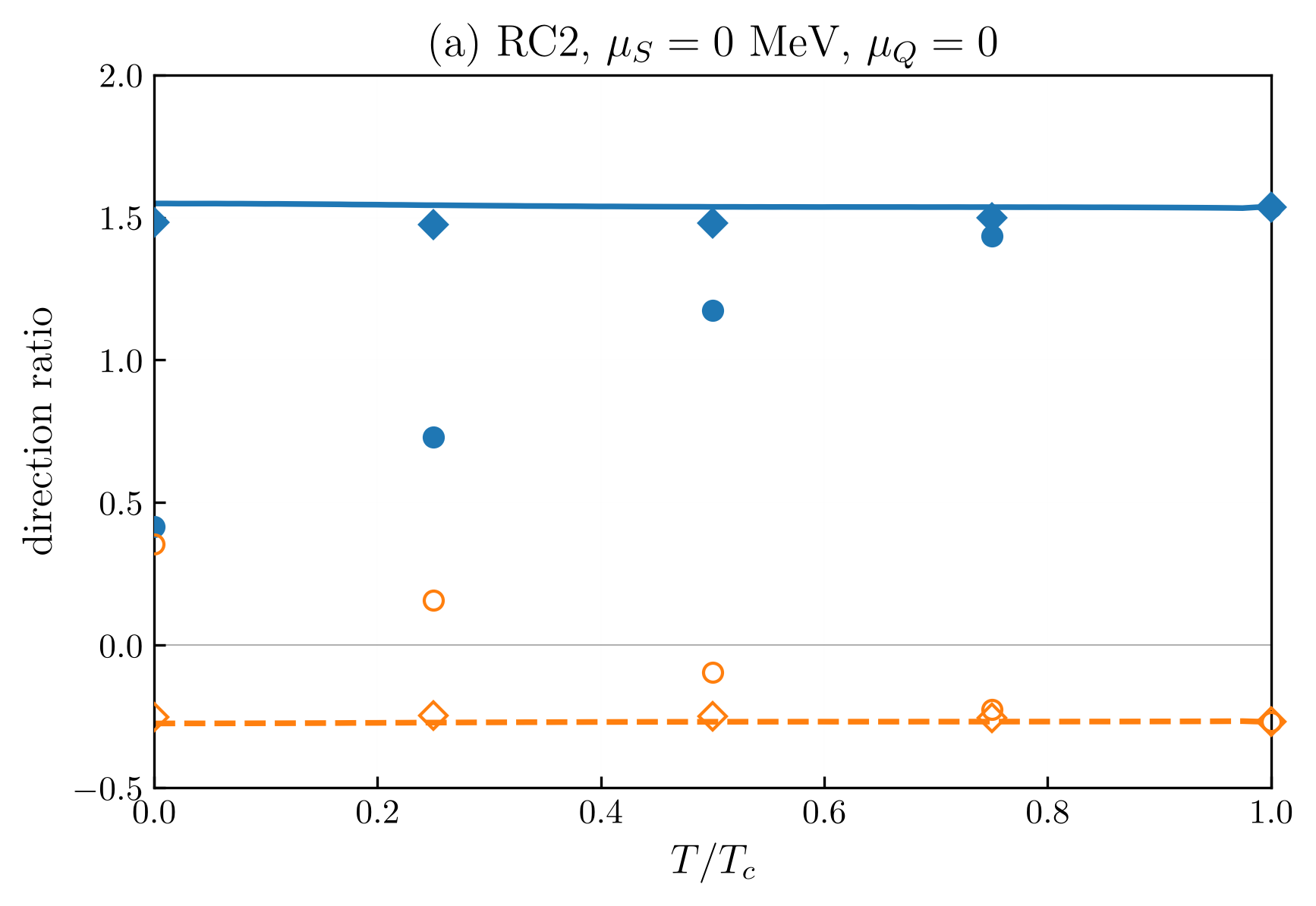}
    \includegraphics[width=0.45\linewidth]{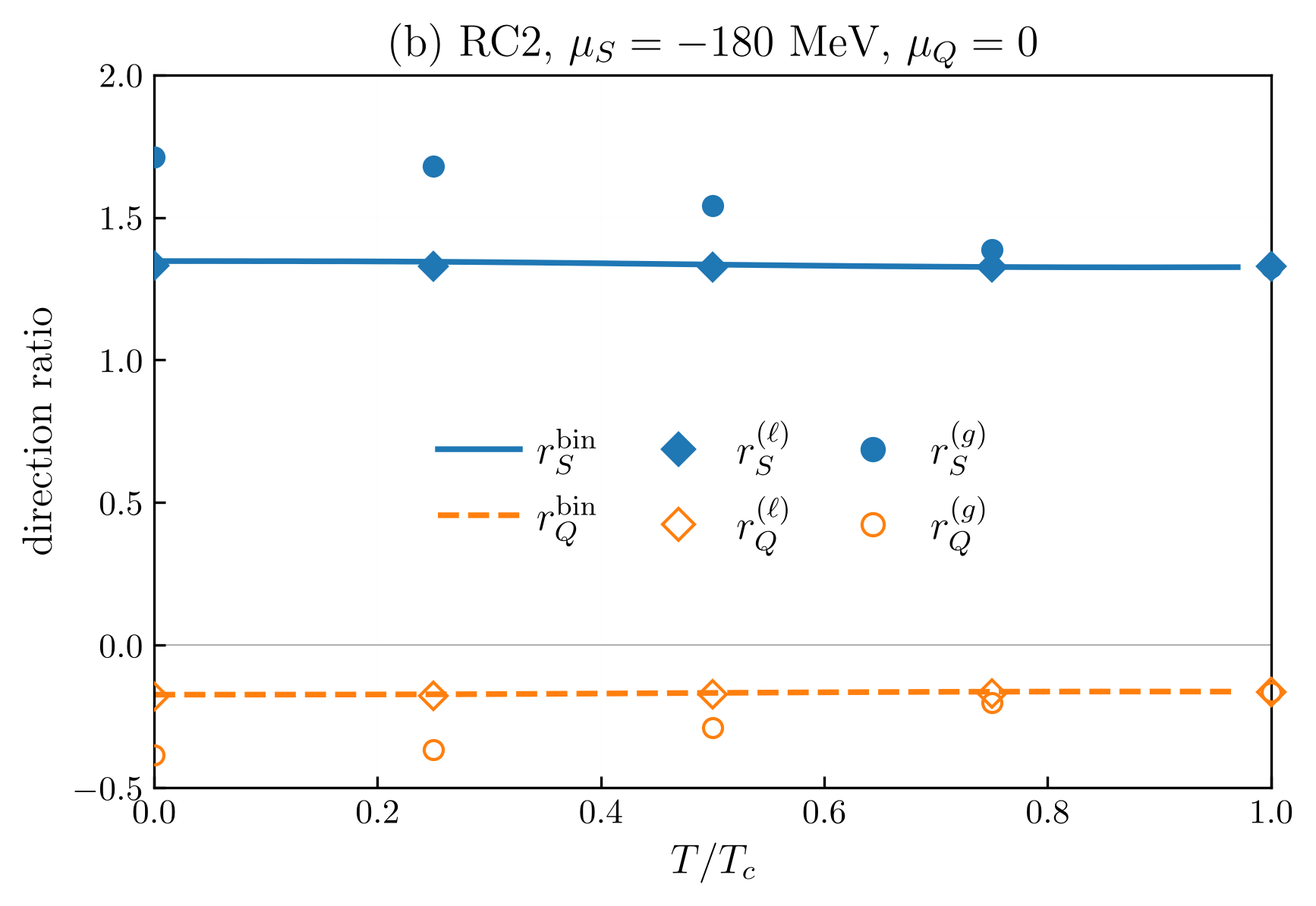}
    \caption[Order-parameter directions for the strangeness-dominated transition.]{Direction of the density discontinuity along the binodal and of the soft eigenvectors at the two spinodal boundaries for the strangeness-dominated transition at $\mu_S=0$ (left) and $\mu_S=-180$~MeV (right). The strange and charge components are normalized by the baryon component and shown as functions of $T/T_c$. The convergence of the three directions as $T\rightarrow T_c^-$ identifies the limiting direction of the critical order parameter. Results are shown for RC2 at $\mu_Q=0$.}
    \label{fig:str_op}
\end{figure*}

The direction ratios defined in \Cref{eq:binodal_direction_ratios} and~\Cref{eq:spinodal_direction_ratios} are shown in \Cref{fig:str_op}. For $T/T_c\lesssim0.9$, $r_S^{\rm bin}$ changes by less than $1\%$ at $\mu_S=0$ and by less than $2\%$ at $\mu_S=-180$~MeV. By contrast, both $\Delta n_B$ and $-\Delta n_S$ decrease by approximately a factor of three by $T\simeq0.95T_c$. The transition therefore weakens substantially as it approaches the critical point, while retaining an almost fixed orientation in conserved-charge space. The values $r_S^{\rm bin}\simeq1.5$ at $\mu_S=0$ and $r_S^{\rm bin}\simeq1.3$ at $\mu_S=-180$~MeV show that the change in strangeness density exceeds the corresponding change in baryon density. The exact isoscalar relation for the binodal then gives a negative charge component.

The dense-side soft direction follows the binodal direction closely, differing in $r_S$ by at most approximately $0.07$ at $\mu_S=0$ and $0.02$ at $\mu_S=-180$~MeV. The dilute-side soft direction rotates much more strongly. At $\mu_S=0$, for example, $r_S^{(g)}$ rises from about $0.42$ at $T=0$ toward the binodal value near $1.5$. In both cases, the dilute- and dense-side soft directions converge with the binodal direction as $T\rightarrow T_c^-$. Their convergence provides a direct numerical check that the critical point is governed by a single soft mode and fixes the direction of the order parameter at $T_c$.

\begin{figure}[t]
    \centering
    \includegraphics[width=0.95\linewidth]{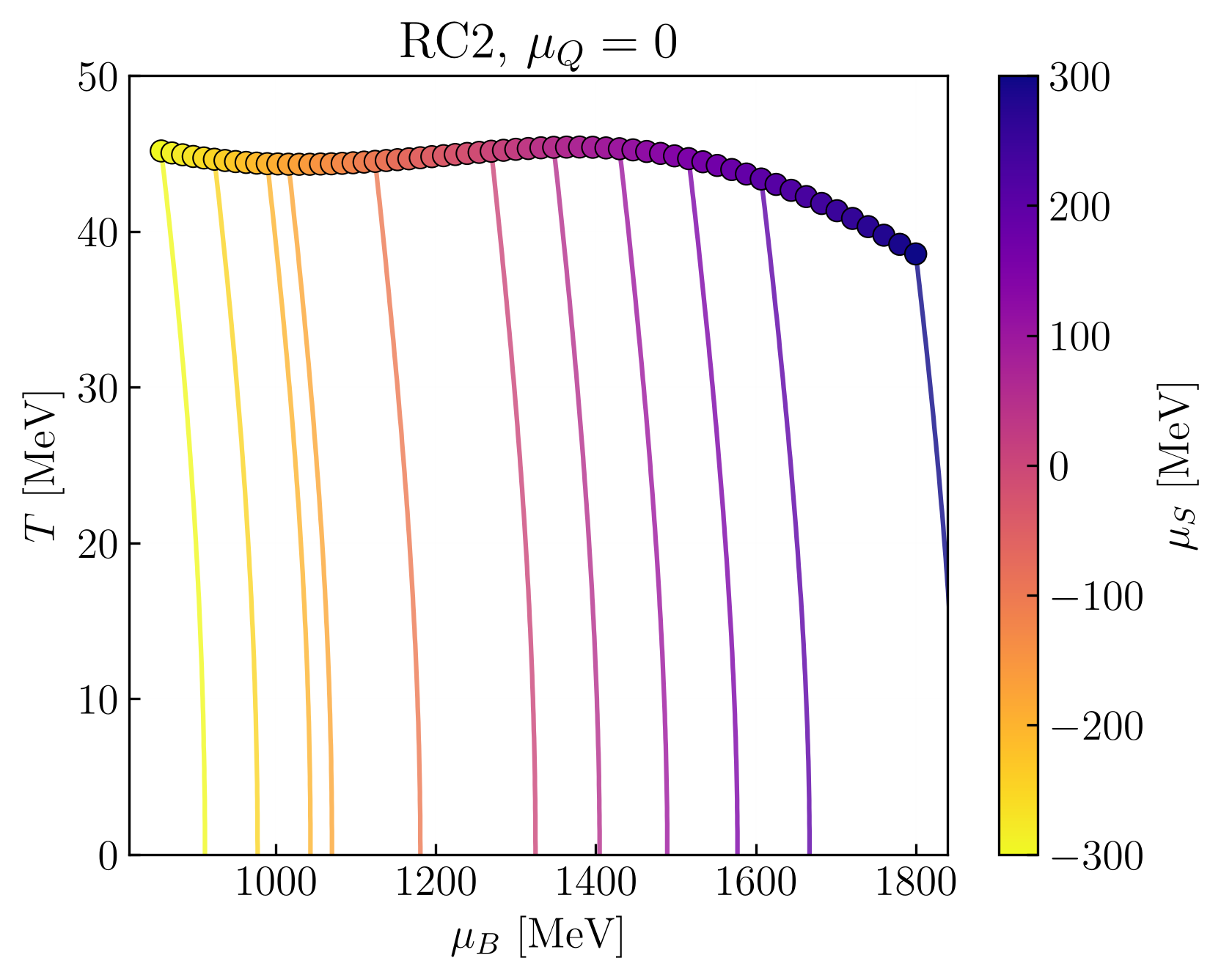}
    \caption{Coexistence lines of the strangeness-dominated transition in the $(\mu_B,T)$ plane for RC2 at $\mu_Q=0$. The color denotes $\mu_S$, and the circles mark the critical points.}
    \label{fig:critical_loci_STR}
\end{figure}

\Cref{fig:critical_loci_STR} shows the evolution of the coexistence lines and their critical endpoints in the $(T,\mu_B)$ plane over $\mu_S\in[-300,300]$~MeV. The critical baryon chemical potential increases nearly monotonically with $\mu_S$ and changes by almost $1$~GeV across this interval. In contrast, $T_c$ remains close to $45$~MeV for negative and small positive values of $\mu_S$, decreasing more clearly only at larger positive $\mu_S$. The transition therefore moves mainly along the chemical-potential direction.

The large, nearly monotonic displacement follows directly from the total strangeness component of the density discontinuity. As quantified below, $-\Delta n_S/\Delta n_B$ remains larger than unity throughout the critical locus, whereas the thermal contribution to the generalized Clausius--Clapeyron slope is comparatively small. The composition term therefore controls the motion of $\mu_{B,c}$ with $\mu_S$.

\begin{figure}[t]
    \centering
    \includegraphics[width=0.99\linewidth]{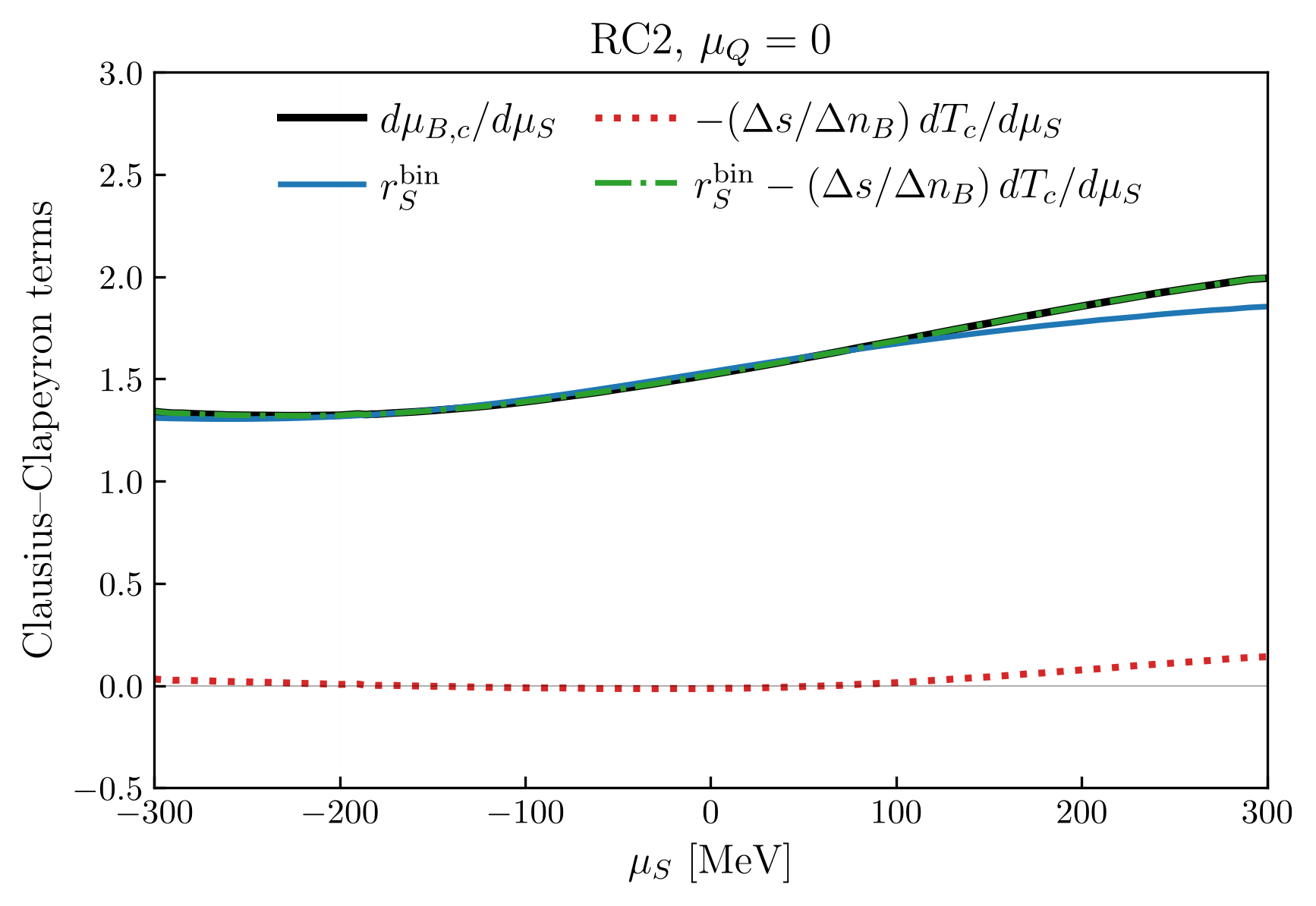}
    \caption{Generalized Clausius--Clapeyron decomposition of the critical locus of the strangeness-dominated transition for RC2 at $\mu_Q=0$. The black curve shows $d\mu_{B,c}/d\mu_S$, while the blue solid and red dotted curves show the composition and thermal terms in \Cref{eq:cc_decomposition}. Their sum is shown by the green dash-dotted curve.}
    \label{fig:str_cc_dec}
\end{figure}

The generalized Clausius--Clapeyron relation provides a direct measure of the critical direction. \Cref{fig:str_cc_dec} shows that \begin{equation}
-\frac{\Delta n_S}{\Delta n_B}>1,
\end{equation}
throughout the complete interval $\mu_S\in[-300,300]$~MeV, increasing from approximately $1.3$ to $1.9$, while the thermal term is comparatively small. The change of strangeness density across the transition is therefore larger than the change of baryon density, which gives a quantitative meaning to the term strangeness dominated. The same density ratio that identifies the critical mode as strangeness dominated therefore also explains the large motion of the critical chemical potential with $\mu_S$.

The strangeness-dominated transition therefore differs qualitatively from the liquid--gas transition. 
It reorganizes octet-dominated matter into decuplet-dominated matter, with $\Delta$ baryons providing most of the baryon-density discontinuity and $\Xi^*$ baryons providing the largest contribution to the strangeness-density discontinuity.
Its order parameter has a large, nearly temperature-independent strange component, and its critical baryon chemical potential changes rapidly with $\mu_S$. Its presence in RC2 and absence in RC4 are consistent with the strong interaction dependence of hyperon-driven instabilities found in relativistic full-octet calculations~\cite{Raduta:2014lja,Torres:2016ydl}. Earlier CMF studies also showed that the high-density phase structure depends strongly on the treatment of the decuplet and its mass generation~\cite{Zschiesche:2000ew}.

The coexistence region in \Cref{fig:sp_bn_str} extends from several times nuclear saturation density to densities above $10n_{\rm sat}$. The strange-baryon and decuplet interactions are poorly constrained in this regime, which makes the quantitative location and even the existence of the transition strongly model dependent. Moreover, $\mu_Q=0$ selects an isospin-symmetric hadronic slice rather than charge-neutral matter in weak equilibrium. Applying the model to neutron star matter would require adding leptons and imposing beta equilibrium and net charge neutrality, which would determine $\mu_Q$ self-consistently. A comprehensive study would also enable the quark sector and compare the pressures of the hadronic and deconfined solutions. If a deconfined phase becomes thermodynamically favored before the hadronic transition, it preempts this transition in equilibrium; the hadronic solutions may then persist as metastable states until their spinodal boundary or may disappear altogether. The present results should therefore be interpreted as a characterization of the hadronic RC2 phase structure rather than as a direct prediction for neutron star matter.

\section{Conclusions}
\label{sec:conclusions}

In this work, we studied how a finite strangeness chemical potential reorganizes the phase structure of the Chiral Mean Field model. By treating $T$, $\mu_B$, and $\mu_S$ as independent thermodynamic variables at $\mu_Q=0$, we followed both the low-density liquid--gas transition and the high-density transition toward strangeness-rich hadronic matter. The analysis combined the coexistence densities, the fixed-temperature $BQS$ susceptibility matrix, and the generalized Clausius--Clapeyron relation. This allowed us to determine not only how the phase boundaries move, but also which conserved-charge densities participate in the corresponding instabilities.

A central outcome of this work is the release of \cmfpp{} \texttt{v2.0.0} as part of the MUSES framework~\cite{MUSES:2023hyz,muses_calculation_engine_2025_v1,pelicer_2026_24091453}. The release provides the finite-temperature CMF thermodynamics used here, the C and RC parametrizations, general vector self-couplings~\cite{Legred:2026cre}, configurable hadronic and quark particle contents, Jacobian-based second-order susceptibilities, and the numerical machinery required to retain and classify multiple mean-field solutions. The code can be used as a stand-alone equation of state module or in the MUSES Calculation Engine. Together with the input files and analysis scripts accompanying this work, it makes the multidimensional phase-structure calculation reproducible and provides a basis for future CMF equation of state tables and systematic parameter studies.

For the liquid--gas transition, a negative $\mu_S$ increases the strange content of both coexisting phases and gives the $BQS$ direction of phase separation a non-negligible strangeness component. The transition nevertheless remains predominantly driven by the baryon density over the range considered here. The CMF RC2 and RC4 parametrizations give a similar description near the range suggested by hypernuclear binding energies, but their predictions differ as $\mu_S$ is lowered and multistrange baryons become increasingly important. 
The Clausius--Clapeyron decomposition shows that this difference is not determined by the density discontinuities alone. 
In RC2, the displacement of the coexistence line is largely controlled by its change in composition, whereas in RC4 the thermal contribution becomes important and produces a substantially non-isothermal trajectory.

The high-density transition in RC2 has a qualitatively different order parameter from the liquid--gas transition. The change in strangeness density is larger than the corresponding change in baryon density, and its normalized direction varies by less than $2\%$ with temperature up to $0.9T_c$. At $\mu_S=0$ and $-180$~MeV, the particle compositions identify a reorganization from octet-dominated to decuplet-dominated matter: $\Delta$ baryons supply most of $\Delta n_B$, while $\Xi^*$ baryons are the largest contributors to $\Delta n_S$. The binodal discontinuity and the soft directions at the two spinodal boundaries approach the same strangeness-dominated direction as the critical point is approached. The strong strange component also explains the large displacement of the critical baryon chemical potential in the $\mu_B$ vs. $T$ plane across the explored $\mu_S$ interval. No corresponding transition was found with RC4, illustrating the sensitivity of this part of the phase diagram to the poorly constrained interactions of strange and decuplet baryons at high density.

The present results were obtained at $\mu_Q=0$ with the quark sector disabled and therefore characterize an isospin-symmetric slice of the hadronic CMF phase diagram. Applying them to neutron stars requires leptons, beta equilibrium, and net charge neutrality, while applications to heavy-ion collisions require the appropriate global charge and strangeness constraints. Nearly isospin-symmetric matter in low-energy heavy-ion collisions may nevertheless provide a setting in which some of the qualitative behavior discussed here is relevant. Future work should determine how the hadronic transitions compete with deconfinement under these system-specific constraints. The present calculation provides a controlled baseline for that extension and clarifies the response of the hadronic CMF model to finite $\mu_S$, which is relevant to heavy-ion matter and may also arise locally in neutron star mergers out of weak equilibrium.

\section*{Acknowledgments}
This research was supported by the NSF under the MUSES collaboration OAC--2103680.
This research used the Delta advanced computing and data resource which is supported by the National Science Foundation (award OAC--2005572), the US-DOE Nuclear Science Grant No. DE-SC0023861,  and the State of Illinois. Delta is a joint effort of the University of Illinois Urbana-Champaign and its National Center for Supercomputing Applications.
This work used Jetstream2 at Indiana University and Open Storage Network at NCSA through allocation PHY230156 from the Advanced Cyberinfrastructure Coordination Ecosystem: Services \& Support (ACCESS) program \citep{NSFACCESS}, which is supported by National Science Foundation grants \#2138259, \#2138286, \#2138307, \#2137603, and \#2138296.
VD additionally acknowledges support from the U.S. Department of Energy under grant DE-SC0024700, from the National Science Foundation under grants NP3M PHY-2116686, PHY-2621752, PHY-2623480, and from the Simons Foundation under award Pivot Fellow-00016159.

\section*{Data availability}
The code associated with this work will be made publicly available at the repository~\cite{pelicer_2026_24091453} soon after journal publication.

\appendix

\section{The baryon decuplet in the mean-field CMF}
\label{app:decuplet}

This appendix expands on the CMF formalism applied to the spin-$3/2$ baryon decuplet. The decuplet flavor wave functions are described by a totally symmetric rank-three flavor tensor $T^{abc}$, where $a,b,c = 1,2,3$ corresponds to $u,d,s$. Each component also carries a Rarita--Schwinger spinor index, which we suppress
\begin{equation}
\begin{alignedat}{2}
T^{111} &= \Delta^{++}, 
&\qquad T^{112} &= \tfrac{1}{\sqrt3}\,\Delta^{+}, \\[2pt]
T^{122} &= \tfrac{1}{\sqrt3}\,\Delta^{0}, 
& T^{222} &= \Delta^{-}, \\[2pt]
T^{113} &= \tfrac{1}{\sqrt3}\,\Sigma^{*+}, 
& T^{123} &= \tfrac{1}{\sqrt6}\,\Sigma^{*0}, \\[2pt]
T^{223} &= \tfrac{1}{\sqrt3}\,\Sigma^{*-}, 
& T^{133} &= \tfrac{1}{\sqrt3}\,\Xi^{*0}, \\[2pt]
T^{233} &= \tfrac{1}{\sqrt3}\,\Xi^{*-}, 
& T^{333} &= \Omega^{-}.
\end{alignedat}
\label{eq:dec:components}
\end{equation}
Each physical state is normalized such that $\sum_{abc}|T^{abc}|^{2}=1$, where the sum runs over all index permutations. The relevant representation product is  $\mathbf{10}\otimes\overline{\mathbf{10}} = \mathbf{1}\oplus\mathbf{8}\oplus\mathbf{27}\oplus\mathbf{64}$.
The adjoint representation appears only once, whereas it appears twice in
$\mathbf{8}\otimes\overline{\mathbf{8}}$. The decuplet therefore has no
independent $F/D$ mixing parameter and no direct analogue of the octet
parameters $\alpha_S$ and $\alpha_V$. Its interaction with a meson nonet is
specified by one octet and one singlet coupling in each channel
\cite{Zschiesche:2000ew}. For the scalar and vector fields,
\begin{equation}
\begin{split}
&\mathcal{L}^{D}=-g_{8}^{D} \bar T^{abc}S_{c}^{~d}T_{abd}
-g_{1}^{D}\, \bar T^{abc}T_{abc}\mathrm{Tr}S \\
&\; -g_{8}^{VD}\bar T^{abc}\gamma_\mu V_{c}^{~d,\mu}T_{abd}
-g_{1}^{VD} \bar T^{abc}\gamma_\mu T_{abc}
\mathrm{Tr}V^\mu .
\end{split}
\label{eq:dec}
\end{equation}
Thus, unlike the octet where the $\Lambda$ and $\Sigma$ can couple differently despite having the same strangeness, the decuplet couplings are fixed by the flavor content of each state.

For any diagonal mean field $M=\mathrm{diag}(M_u,M_d,M_s)$, the symmetry of $T^{abc}$ gives 
\begin{equation}
    \bar T^{abc}M_{c}^{~d}T_{abd}=\frac{1}{3}\left(N_uM_u+N_dM_d+N_sM_s\right)\,\bar T \,T.
\end{equation}
A decuplet baryon therefore couples to the flavor-weighted average of the diagonal mean fields. Here $N_u+N_d+N_s=3$ and $N_u-N_d=2I_3$. We use $N_S \equiv N_s = -S$ for the number of strange valence quarks. The resulting effective masses and chemical potentials retain the form of \Cref{eq:m_eff} and \Cref{eq:mu_eff}, with all couplings fixed by flavor content alone.

We include two additional symmetry-breaking contributions to the in-medium mass of the strange members of the decuplet
\begin{equation}
\begin{split}    
\mathcal{L}^{D}_{\rm Y}
&= -3\,m_{\rm res}\,\bar T^{abc}\mathcal{S}_{c}^{~d}T_{abd} \\
  &\;-\;3\,m_{3}^{D}\,\bar T^{abc}\mathcal{S}_{c}^{~d}T_{abd}\,
   \mathrm{Tr}\bigl(S-S_0\bigr).
\end{split}
\label{eq:dec_sb}
\end{equation}
where $\mathcal{S}=\mathrm{diag}(0,0,1)$ projects 
onto the strange flavor direction.
The parameter $m_{\rm res}$ provides a bare correction associated with the strange-quark content, while $m_3^D$ modifies the in-medium masses and may be adjusted to reproduce a chosen decuplet potential.
The scalar couplings are
\begin{equation}
\begin{aligned}
g_{\sigma i}
&=
\frac{(3-N_S^i)g_8^D}{3\sqrt2}
+\sqrt2\,g_1^D
+\sqrt2\,N_S^i\kappa_i m_3^D,
\\
g_{\zeta i}
&=
\frac{N_S^i g_8^D}{3}
+g_1^D
+N_S^i\kappa_i m_3^D,
\\
g_{\delta i}
&=
\frac{2I_3^i g_8^D}{3\sqrt2}.
\end{aligned}
\label{eq:dec:gS}
\end{equation}
We parameterize the two independent scalar couplings as
\begin{equation}
    g_1^D=\sqrt{2}\,g_{DS}\alpha_{DS},
    \qquad
    g_8^D=3\sqrt{2}\,g_{DS}(1-\alpha_{DS}).
\end{equation}
Here, $\alpha_{DS}$ parameterizes the relative singlet--octet
strength and is not an $F/D$ mixing parameter. We use the values $g_{DS}=-4.106$ and $\alpha_{DS}= 0.351$, which correspond to $g_1^D= -2.038$ and $g_8^D=-11.306$. The corresponding vector couplings are
\begin{equation}
\begin{aligned}
g_{\omega i}
&=
\frac{(3-N_S^i)g_8^{VD}}{3\sqrt2}
+\sqrt2\,g_1^{VD},
\\
g_{\phi i}
&=
-\frac{N_S^i g_8^{VD}}{3}
-g_1^{VD},
\\
g_{\rho i}
&=
\frac{2I_3^i g_8^{VD}}{3\sqrt2}.
\end{aligned}
\label{eq:dec:gV}
\end{equation}
The corresponding bare mass term is
\begin{equation}
\Delta m_i=N_S^{i}\bigl(m_{\rm res}-\kappa_i\,m_3^{D}\,(\sqrt{2} \sigma_0 + \zeta_0) \bigr), \qquad \Delta m_\Delta=0.
\label{eq:dec:deltam}
\end{equation}
The vacuum contribution from $m_3^D$  cancels the corresponding contribution from the scalar couplings in the vacuum in \Cref{eq:dec:gS}. The phenomenological factor $\kappa_i$ is used in the octet and decuplet sectors to adjust the contribution to the $\Xi$, $\Xi^\ast$ and $\Omega$ optical potentials. 

The vector couplings follow the same flavor structure as the scalar ones,
under $g_8^D\to g_8^{VD}$ and $(\sigma,\zeta,\delta)\to(\omega,\phi,\rho)$. We parameterize their overall strength 
\begin{equation}\label{eq:dec:vd}
    V_\Delta= \frac{g_{\omega \Delta}}{g_{\omega N}}
\end{equation}
with the $\mathrm{SU}(6)$ choice corresponding to $V_\Delta=1$. Once $V_\Delta$ is chosen, the remaining decuplet vector couplings follow from the flavor structure summarized in \Cref{tab:dec} (see also~\cite{Weissenborn:2011ut}).

\begin{table*}[t]
\centering
\caption{Decuplet valence content and couplings. Within each multiplet $k=0,\dots,3-N_S$ labels the charge states, so that $N_u-N_d=2I_3$; every
member has spin degeneracy $\gamma_i=4$. The chiral parts of $g_{\sigma i}$ and $g_{\zeta i}$ are shown; the explicit-breaking additions of Eq.~\eqref{eq:dec_sb} are omitted.}
\label{tab:dec}
\begin{tabular}{lcccccccc}
\toprule
& $N_u$ & $N_d$ & $N_S$ & $I$
& $g_{\sigma i}/g_{DS}$ & $g_{\zeta i}/(\sqrt2\,g_{DS})$
& $g_{\omega i}/(V_\Delta g_{\omega N})$ & $g_{\phi i}/(V_\Delta g_{\omega N})$\\
\midrule
$\Delta$     & $3-k$ & $k$ & 0 & $3/2$ & $3-\alpha_{DS}$ & $\alpha_{DS}$
             & $1$   & $0$\\
$\Sigma^{*}$ & $2-k$ & $k$ & 1 & $1$   & $2$ & $1$
             & $2/3$ & $-\sqrt2/3$\\
$\Xi^{*}$    & $1-k$ & $k$ & 2 & $1/2$ & $1+\alpha_{DS}$ & $2-\alpha_{DS}$
             & $1/3$ & $-2\sqrt2/3$\\
$\Omega$     & $0$   & $0$ & 3 & $0$   & $2\alpha_{DS}$ & $3-2\alpha_{DS}$
             & $0$   & $-\sqrt2$\\
\bottomrule
\end{tabular}
\end{table*}

Since $\mathcal{S}=\tfrac13 \mathbbm{1}-\tfrac{1}{\sqrt3}\lambda_8$ satisfies $\mathrm{Tr}(\mathcal{S}\lambda_a)=0$ for $a=1,\dots,7$, the breaking lies entirely along the hypercharge and singlet directions and leaves the pion and kaon PCAC relations intact, as for the octet term~\cite{Papazoglou:1998vr}. 
 
The octet and singlet contributions in \Cref{eq:dec:gS} and~\Cref{eq:dec:gV} are linear in the strange valence content. The phenomenological factor $\kappa_i$ and the explicit symmetry-breaking contribution in \Cref{eq:dec:deltam}, however, need not preserve an exact in-medium equal spacing relation. We therefore use the complete species-dependent couplings when calculating the decuplet masses and potential.

\bibliography{inspire,not_inspire}

\end{document}